

\let\AMSfonts1 
\magnification 1200
\baselineskip=12pt
\hsize=16.5truecm \vsize=23 truecm \voffset=.4truecm
\parskip=14 pt

\def\idty{{\leavevmode{\rm 1\ifmmode\mkern -5.4mu\else\kern -.3em\fi I}}}
\def\Ibb #1{ {\rm I\ifmmode\mkern -3.6mu\else\kern -.2em\fi#1}}
\def\ibb #1{\leavevmode\hbox{\kern.3em\vrule
     height 1.5ex depth -.1ex width .2pt\kern-.3em\rm#1}}
\def\Nl{{\Ibb N}} \def\Cx {{\ibb C}} \def\Rl {{\Ibb R}}
\def\lessblank{\parskip=5pt \abovedisplayskip=2pt
          \belowdisplayskip=2pt }
\def\eproclaim{\par\endgroup\vskip0pt plus100pt\noindent}
\def\proof#1{\par\noindent {\bf Proof #1}\          
         \begingroup\lessblank\parindent=0pt}
\def\QED {\hfill\endgroup\break
     \line{\hfill{\vrule height 1.8ex width 1.8ex }\quad}
      \vskip 0pt plus100pt}
\def\Examp#1. #2:{\par\noindent{\bf #1. #2}\hfill\break}

\def\Bar{\overline}
\def\abs #1{{\left\vert#1\right\vert}}
\def\bra #1>{\langle #1\rangle}
\def\bracks #1{\lbrack #1\rbrack}

\def\id{\mathop{\rm id}\nolimits}

\def\norm #1{\left\Vert #1\right\Vert}
\def\Norm#1#2{\mathopen#1\Vert{#2}\mathclose#1\Vert}
\def\set #1{\left\lbrace#1\right\rbrace}
\def\stt{\,\vrule\ }
\def\Set#1#2{#1\lbrace#2#1\rbrace}  
\def\tr {\mathop{\rm tr}\nolimits}

\def\lbk{{\vcenter{\vbox{\hrule height .6pt
                   \hbox{\vrule width 1.2pt  height 10pt \kern 3pt }
                   \hrule height .6pt}}}}
\def\rbk{{\vcenter{\vbox{\hrule height .6pt
                   \hbox{\kern 3pt  \vrule width 1.2pt  height 10pt }
                   \hrule height .6pt}}}}
\def\bracks#1{\mkern5mu\lbk{#1}\rbk\mkern5mu}
\def\Pbrack#1{\big\lbrace\mkern-9.8mu\big\lbrace {#1}
              \big\rbrace\mkern-9.8mu\big\rbrace} 
\def\Bra#1{\langle#1\vert}
\def\Ket#1{\vert#1\rangle}

\def\limsup{\mathop{\hbox{$\overline{\rm lim}$}}}

\def\phi{\varphi}
\def\epsilon{\varepsilon}
\def\3{\ss}
\def\Re{\mathchar"023C\mkern-2mu e} 
\def\Im{\mathchar"023D\mkern-2mu m} 
\def\ie{i.e.,\ }
\def\eg{e.g.,\ }
\def\QM{quantum mechanics}
\def\hx{\widehat x}
\def\hp{\widehat p}
\font\mbf=cmmib10
\font\mbbf=cmmib10 scaled \magstep1
\font\mbbbf=cmmib10 scaled \magstep2
\def\fathb{\hbox{\bf\raise 3pt\hbox{--}\kern-6pt\mbf h}}
\def\fathbb{\hbox{\bf\raise 4pt\hbox{--}\kern-6pt\mbbf h}}
\def\bigfatjhh{$\hbox{\mbbbf j}_{\textstyle\fathbb\fathbb'}$}
\ifx\AMSfonts1

\fi
\def\hb{\hbar}
\def\h{_\hb}
\def\oprime#1{{\def\hb{{\hbar'}}#1}}

\def\hbzero#1{{\def\hb{0}#1}}

\def\nohb#1{{\def\hb{}#1}}
\def\Weylo{E_\hb}     
\def\Weylg{W^\hb}     
\def\Ps{\Xi}          
\def\pstr{\alpha^\hb} 
\def\psm#1#2{{d#1\,d#2\over(2\pi\hb)^d}\ }  
\def\tev#1#2{\gamma^{#1}_{#2}}  
\def\mud{\mu_d}             
\def\syf#1#2{\sigma(#1,#2)} 
\def\syfdots{\sigma}        
\def\parity{\Pi}

\ifx\AMSfonts1
\global\def\A{\hbox{\frak A}}%
\global\def\LagrM{\hbox{\frak L}}
\global\def\Wigner{\hbox{\frak W}\h}%
\global\def\jWigner{\jt^{\hbox{\frakito W}}}%
\else
\global\def\A{{\cal A}}

\global\def\LagrM{{\cal L}}
\global\def\Wigner{{\cal W}\h}
\global\def\jWigner{\jt^{\cal W}}
\fi
\def\B{{\cal B}}

\def\dS{{\rm d}S}
\def\coh{\chi\h}   
\def\Coh{\Gamma\h} 
\def\Hosc{H^{\rm osc}\h}
\def\mc{{\bf m}}  
\def\L#1{{\cal L}^{#1}}  
\def\C#1{{\cal C}^{#1}}  
\def\H{{\cal H}}
\def\j#1#2{j_{#1#2}}
\def\jhh{\j\hb{\hb'}}
\def\jt{\hbox{$\j{}{}$}}   
\def\jj#1#2#3{\j{#1}{#2}\j{#2}{#3}}
\def\jlim{\jt\mkern-2mu\hbox{-}\mkern-5mu\lim}
\def\jslim{\jt^*\mkern-2mu\hbox{-}\mkern-5mu\lim}
\def\CAj{\C{}(\A,\jt)}
\def\tj{\widetilde\jmath}  
\def\jconv{\jt-convergent}
\def\jsconv{\jt*-convergent}      
\def\mfat#1{{\def\jt{\hbox{\mbf j}}\let\hb\fathb\bf#1}}
\def\hbs{$\hb$-sequence}

\catcode`@=11
\def\ifundefined#1{\expandafter\ifx\csname
                        \expandafter\eat\string#1\endcsname\relax}
\def\atdef#1{\expandafter\def\csname #1\endcsname}
\def\atedef#1{\expandafter\edef\csname #1\endcsname}
\def\atname#1{\csname #1\endcsname}
\def\ifempty#1{\ifx\@mp#1\@mp}
\def\ifatundef#1#2#3{\expandafter\ifx\csname#1\endcsname\relax
                                  #2\else#3\fi}
\def\eat#1{}
\newcount\refno \refno=1
\def\labref #1 #2 #3\par{\atdef{R@#2}{#1}}
\def\lstref #1 #2 #3\par{\atedef{R@#2}{\number\refno}
                              \advance\refno by1}
\def\txtref #1 #2 #3\par{\atdef{R@#2}{\number\refno
      \global\atedef{R@#2}{\number\refno}\global\advance\refno by1}}
\def\doref  #1 #2 #3\par{{\refno=0
     \vbox {\everyref \item {\reflistitem{\atname{R@#2}}}
            {\d@more#3\more\@ut\par}\par}}\refskip }
\def\d@more #1\more#2\par
   {{#1\more}\ifx\@ut#2\else\d@more#2\par\fi}
\def\@cite #1,#2\@ver
   {\eachcite{#1}\ifx\@ut#2\else,\hskip0pt\@cite#2\@ver\fi}
\def\cite#1{\citeform{\@cite#1,\@ut\@ver}}
\def\eachcite#1{\ifatundef{R@#1}{{\tt#1??}}{\atname{R@#1}}}
\def\defonereftag#1=#2,{\atdef{R@#1}{#2}}
\def\defreftags#1, {\ifx\relax#1\relax \let\next\relax \else
           \expandafter\defonereftag#1,\let\next\defreftags\fi\next }
\def\@utfirst #1,#2\@ver
   {\author#1,\ifx\@ut#2\afteraut\else\@utsecond#2\@ver\fi}
\def\@utsecond #1,#2\@ver
   {\ifx\@ut#2\andone\author#1,\afterauts\else
      ,\author#1,\@utmore#2\@ver\fi}
\def\@utmore #1,#2\@ver
   {\ifx\@ut#2\and\author#1,\afterauts\else
      ,\author#1,\@utmore#2\@ver\fi}
\def\authors#1{\@utfirst#1,\@ut\@ver}
\catcode`@=12
\let\REF\labref
\def\citeform#1{{\bf\lbrack#1\rbrack}}  
\let\reflistitem\citeform               
\let\everyref\relax                     
\let\more\relax                         
\def\refskip{\vskip 10pt plus 2pt}      
\def\colbr{\hskip.3em plus.3em\penalty-100}  
\def\combr{\hskip.3em plus4em\penalty-100}   
\def\refsecpars{\emergencystretch=50 pt      
                 \hyphenpenalty=100}
\def\Bref#1 "#2"#3\more{\authors{#1}:\colbr {\it #2},\combr #3\more}
\def\Gref#1 "#2"#3\more{\authors{#1}\ifempty{#2}\else:\colbr``#2''\fi
                        ,\combr#3\more}
\def\Jref#1 "#2"#3\more{\authors{#1}:\colbr``#2'',\combr \Jn#3\more}
\def\inPr#1 "#2"#3\more{in: \authors{\eds#1}:\colbr
                          ``{\it #2}'',\combr #3\more}
\def\Jn #1 @#2(#3)#4\more{\hbox{\it#1}\ {\bf#2}(#3)#4\more}
\def\author#1. #2,{\hbox{#1.~#2}}            
\def\sameauthor#1{\leavevmode$\underline{\hbox to 25pt{}}$}
\def\and{, and}   \def\andone{ and}          
\def\noinitial#1{\ignorespaces}
\let\afteraut\relax
\let\afterauts\relax
\def\etal{\def\afteraut{, et.al.}\let\afterauts\afteraut
           \let\and,}
\def\eds{\def\afteraut{(ed.)}\def\afterauts{(eds.)}}
\catcode`@=11
\newcount\eqNo \eqNo=0
\def\lasteq{\secNo.\number\eqNo}
\def\deq#1(#2){{\ifempty{#1}\global\advance\eqNo by1
       \edef\n@@{\lasteq}\else\edef\n@@{#1}\fi
       \ifempty{#2}\else\global\atedef{E@#2}{\n@@}\fi\n@@}}
\def\eq#1(#2){\edef\n@@{#1}\ifempty{#2}\else
       \ifatundef{E@#2}{\global\atedef{E@#2}{#1}}%
                       {\edef\n@@{\atname{E@#2}}}\fi
       {\rm($\n@@$)}}
\def\deqno#1(#2){\eqno(\deq#1(#2))}
\def\deqal#1(#2){(\deq#1(#2))}
\def\eqback#1{{(\advance\eqNo by -#1 \lasteq)}}

\def\eqgroup(#1){{\global\advance\eqNo by1
       \edef\n@@{\lasteq}\global\atedef{E@#1}{\n@@}}}
\outer\def\iproclaim#1/#2/#3. {\vskip0pt plus50pt \par\noindent
     {\bf\dpcl#1/#2/ #3.\ }\begingroup \interlinepenalty=250\lessblank\sl}
\newcount\pcNo  \pcNo=0
\def\lastpc{\number\pcNo} 
\def\dpcl#1/#2/{\ifempty{#1}\global\advance\pcNo by1
       \edef\n@@{\lastpc}\else\edef\n@@{#1}\fi
       \ifempty{#2}\else\global\atedef{P@#2}{\n@@}\fi\n@@}
\def\pcl#1/#2/{\edef\n@@{#1}%
       \ifempty{#2}\else
       \ifatundef{P@#2}{\global\atedef{P@#2}{#1}}%
                       {\edef\n@@{\atname{P@#2}}}\fi
       \n@@}
\def\Def#1/#2/{Definition~\pcl#1/#2/}
\def\Thm#1/#2/{Theorem~\pcl#1/#2/}
\def\Lem#1/#2/{Lemma~\pcl#1/#2/}
\def\Prp#1/#2/{Proposition~\pcl#1/#2/}
\def\Cor#1/#2/{Corollary~\pcl#1/#2/}
\def\Exa#1/#2/{Example~\pcl#1/#2/}
\font\sectfont=cmbx10 scaled \magstep2
\def\bgsecti@n #1. #2\e@h{\def\secNo{#1}\eqNo=0}
\def\bgssecti@n#1. #2\e@h{}
\def\secNo{00}
\def\lookahead#1#2{\vskip\z@ plus#1\penalty-250
  \vskip\z@ plus-#1\bigskip\vskip\parskip
  {#2}\nobreak\smallskip\noindent}
\def\secthead#1. #2\e@h{\leftline{\sectfont
                        \ifx\n@#1\n@\else#1.\ \fi#2}}
\def\bgsection#1. #2\par{\bgsecti@n#1. #2\e@h
        \lookahead{.3\vsize}{\secthead#1. #2\e@h}}
\def\bgssection#1. #2\par{\bgssecti@n#1. #2\e@h
        \lookahead{.3\vsize}{\leftline{\bf#1. #2}}}
\def\bgsections#1. #2\bgssection#3. #4\par{%
        \bgsecti@n#1. #2\e@h\bgssecti@n#3. #4\e@h
        \lookahead{.3\vsize}{\vtop{\secthead#1. #2\e@h\vskip10pt
                             \leftline{\bf#3. #4}}}}
\def\Acknow#1\par{\ifx\REF\doref
     \bgsection. Acknowledgements\par#1\refsecpars
     \bgsection. References\par\fi}
\catcode`@=12
\def\Sec#1{Section~{#1}}
\defreftags Albeverio=AHK, AlbevBook=AFHL, Arai=Ara, Berry=BB,
Belliss=BV, Sirugue=BCSS, Bopp=Bop, PWI=BW, Bruer=Bru, BurdHJ=BS,
Cartwright=Car, Chung=Chu, Daubechies=Dau, Davies=Dav, Esposti=DGI,
Duclos=DH, Nick=Duf, LMD=DW, Emch=Em1, Emchbook=Em2,
Hajo=FLM, Froman=Fr{\accent "7F o}, FGH=GW, Grossman=Gro,
Hagedorn=Hag, Helffer=Hel, Hepp=Hep, Holevo=Hol, Hormander=Hor,
Hudson=Hud, Kato=Kat, Boch=Kat, Landsman=Lan, Liebspin=Lie,
Liebcoh=LS, Maslov=Mas, Narcowich=Nar, Omnes=Omn, IMF=RW,
Rieffel=Ri1, RieffAMS=Ri2, Rieffprep=Ri3, Robert=Rob, Schiff=Sch,
Simon=Sim, Takahashi=Tak, Aubrey=Tru, Voros=Vor, QHA=We1, PHU=We2,
KAC=We3, CLJ=We4, CLD=We5, CLN=We6, IHQ=WW, Wigner=Wig, Wresin=WS, ,

\line{}
\vskip 2.0cm
\font\BF=cmbx10 scaled \magstep 3
{\BF \baselineskip=25pt
\centerline{ The Classical Limit }
\centerline{ of Quantum Theory}
}
\vskip 2.0cm
\centerline{
\bf R.F. Werner
\footnote{$^1$}%
{{\sl
FB Physik, Universit\"at Osnabr\"uck, 49069 Osnabr\"uck, Germany
}}%
$^,$\footnote{$^{2}$}
{{ \sl Electronic mail:\quad
   \tt reinwer@dosuni1.rz.Uni-Osnabrueck.DE}}
}
\vskip 1.0cm

{\narrower\narrower\noindent
{\bf Abstract.}\
For a quantum observable $A_\hbar$ depending on a parameter $\hbar$
we define the notion ``$A_\hbar$ converges in the classical limit''.
The limit is a function on phase space. Convergence is in norm in
the sense that $A_\hbar\to0$ is equivalent with
$\Vert A_\hbar\Vert\to0$. The $\hbar$-wise product of convergent
observables converges to the product of the limiting phase space
functions. $\hbar^{-1}$ times the commutator of suitable observables
converges to the Poisson bracket of the limits. For a large class of
convergent Hamiltonians the $\hbar$-wise action of the corresponding
dynamics converges to the classical Hamiltonian dynamics. The
connections with earlier approaches, based on the WKB method, or on
Wigner distribution functions, or on the limits of coherent states
are reviewed.
\par}
\vskip20 pt

\noindent {\bf Physics and Astronomy classification scheme PACS
(1994):}\hfill\break
03.65.Sq, 
03.65.Db  
\vfil\eject

\bgsection 1. Introduction

The problem of taking the limit  of \QM\ as $\hbar\to0$ is as old as
\QM\ itself. Indeed, under the name ``correspondence principle'' it
was one of the important guidelines for the construction of the
theory itself. Naturally, there is a vast literature on the subject,
and it requires some justification to add yet another paper to it. I
will therefore begin by stating the aims of the present paper more
carefully than usual, and proceed to review some of the existing
approaches to the classical limit with regard to these aims. This
will be done in a separate subsection of the introduction.
In \Sec2 and \Sec3 we describe the basic notions of our approach.
It is based on a set of ``comparison maps'' $\jhh$ which relate
observables at different values of  $\hbar$. This framework was
originally designed for applications in statistical mechanics
\cite{KAC}, and has many further conceivable applications. In
\Sec2 it is shown that this furnishes a language in which the
convergence of sequences of observables, and the theorems of the
desired type can be adequately expressed. The definition of the
comparison maps $\jhh$ requires some additional structure from phase
space quantum mechanics, and is undertaken in \Sec3. \Sec4
gives an extensive list of examples and applications. We hope that
this section especially will help to convince the reader that the
present approach to the classical limit is a natural, if not
canonical one. \Sec5 contains the more technical aspects, including,
of course, the proofs of the main results. Some of these technical
points, notably the proofs of the theorems about convergence of
commutators to Poisson brackets, and the convergence of dynamics
were beyond the scope of a single journal article, and will
therefore be treated in a separate publication \cite{CLD}. The
concluding \Sec6 contains previews of such further extensions, and
also some remarks about how some simplifying assumptions (like the
boundedness of Hamiltonians) can be relaxed.

\bgssection 1.1. Motivation and review of the literature

There are basically two reasons for studying the classical limit.
The first is concerned with the architecture of
theoretical physics, and demands the reconstruction of classical
mechanics in terms of its supposedly more comprehensive successor.
This ``correspondence principle'' was part of the supporting
evidence for the new quantum theory. Now that this is hardly needed
anymore, some theorists feel that there is no more reason to study
the classical limit.  Some physicists also seem to feel uneasy about
the sacrilege of changing the value of the Fundamental Constant
$\hbar=1.0545887*10^{-34}\,kg\,m^2/s$ (or $\hb=1$ in more practical
units). Are we free to do this without talking about a different
possible world of no relevance to our own? This leads to the second
motivation for discussing the classical limit: it is seen mainly as
a practical tool for the simplified approximate evaluation of
quantum mechanical predictions. In this interpretation a limit
theorem says that the classical treatment is accurate (within certain
bounds) as long as the relevant observables change sufficiently
slowly relative to the phase space scale fixed by $\hb$. The
introduction of a changeable parameter $\hb$ is then merely a
convenient shorthand for this comparison. What makes it especially
convenient is that the comparison parameter $\hb$ will show up in
all those places, where we are used to seeing the constant $\hb$ in
the textbooks.

For the mathematical formulation of the classical limit both
readings amount to the same thing. The following are some of the
features, which one might ask of a satisfactory explanation, and
which the present paper aims to implement.

{\parskip=5pt
\item{(a)}
The limit should be defined for the {\it whole theory}, not of
certain isolated aspects. That is, we should define the limits of
general states, observables, and expectation values, and these
should go to their classical counterparts.
\item{(b)}
The definition should be conceptually {\it simple} and {\it
general}. That is, it should be appropriate for inclusion in a basic
course on quantum mechanics. It should not depend on the choice of a
special (\eg  quadratic or classically integrable) Hamiltonian, or
special (\eg coherent) states.
\item{(c)}
It should be a rigorous version of accepted {\it folklore} on the
subject. For example, the limit of $-\hbar^2/(2m)\Delta +V(x)$
should be the Hamiltonian function $p^2/2m+V(q)$, and some intuition
should be given, for what kinds of observables the classical
approximation is sensible.
\item{(d)}
The limit should be in the {\it strongest topology} possible.
We want the statement of the limit to be a equivalent to an
asymptotic estimate of {\it operator norms} for observables and trace
norms for states. These norms carry special significance in the
statistical interpretation of quantum theory, since they correspond
to uniform estimates on probabilities.
\item{(e)}
In the limit, the {\it product} of bounded operators should become
the product of functions on phase space.
\item{(f)}
In the limit, ``$i/\hbar$ times a commutator'' should
become the {\it Poisson bracket} of the limits.
\item{(g)}
The quantum mechanical {\it time evolution} should converge
(uniformly in finite time intervals) to the classical Hamiltonian
evolution.
\item{(h)}
{\it Equilibrium states} (canonical Gibbs states) and {\it partition
functions} of quantum theory should converge to their classical
counterparts.
\item{}}

\noindent
On the other hand, we can distinguish in the literature the
following approaches to the classical limit, each of which naturally
has a considerable overlap of results and applications with the
approach we are going to present. This list is necessarily
incomplete, and no attempt has been made to evaluate the historical
development of the subject, or to decide any priority claims. Nor
can we adequately portray the merits of the different schools since
our perspective is limited to the comparison with the approach of
the present paper.

\item{(A)}
{\it The WKB method.}
\cite{Maslov,Schiff,Helffer,Froman,Duclos,BurdHJ}
One virtue of this well-known approach is that it is so
close to Schr\"odinger's beautiful series of papers establishing his
wave mechanics. It fails mainly on item (a): the Schr\"odinger
equation is only one aspect of \QM, and its short wave asymptotics is
only one aspect of the classical limit. For example, it seems hopeless
to try to understand the operator properties (e) and (f) in WKB terms.
The WKB wave functions do correspond to (a subclass of ) convergent
states in our approach (see \Sec4.8). Their limits are measures
supported by Lagrangian manifolds in phase space, hence they have a
curious intermediate position between point measures and general
measures.

\item{(B)} {\it Wigner functions.}
\cite{Wigner,Berry,Bruer,Sirugue,Arai}
It is often claimed that \QM\ has an equivalent reformulation in
terms of Wigner's phase space distribution functions. The classical
limit could then be stated very simply in terms of these functions.
However, the premise is only partly correct. Since the Wigner
function of a state need not be integrable, it often represents a
``probability'' density, in which an infinite positive probability
is cancelled by an infinite negative probability to give formally
the normalization to unity. This is highly unsatisfactory from the
conceptual point of view. Technically it means that operator norms
(see (d) above) cannot be estimated without artificial smoothness
assumptions \cite{Daubechies}. It is well-known that by averaging
Wigner functions with a suitable Gaussian \cite{Bopp,Cartwright}
these difficulties disappear \cite{Davies,Holevo,QHA}. Moreover, the
Gaussians can be chosen such that in the classical limit this
smearing out becomes negligible anyhow. In their averaged form
Wigner functions play an important role in our approach. For a
discussion of states that have positive Wigner functions ``all the
way to the classical limit'' see \Sec4.10.

\item{(C)} {\it Pseudodifferential and Fourier integral operators.}
\cite{Robert,Voros,Omnes}
Such operators have a rich mathematical theory, whose applications
are by no means confined to the classical limit. However, much of
the rigorous work on the classical limit has been done under this
heading. The ``symbol'' of a pseudodifferential operator is just its
Wigner function, so much of what has been said under (B) applies.
The main weakness is again the lack of control on operator norms,
and hence of probability estimates, unless additional smoothness
assumptions are introduced. Where such assumptions hold, the results
fit well into the framework of the present paper, too.

\item{(D)}
{\it Feynman integrals.} The basic observation here is that the
phase of the Feynman integrand is stationary precisely for the
classical paths, which therefore give the main contribution to the
propagator. To the extent that the Feynman integral and the method
of stationary phase in infinite dimensional spaces can be given a
mathematical meaning, this observation can be made rigorous
\cite{Aubrey,Albeverio}, and reproduces WKB wave functions. The
shortcomings of this approach are therefore similar to the WKB
approach. It is maybe interesting to note that the propagator itself
does {\it not} have a classical limit in our approach, whereas the
time evolution it implements on observables does (see \Sec4.5).

\item{(E)}
{\it Limits of coherent states.}
In the papers \cite{Hepp,Hagedorn} it is shown that in the limit
$\hbar\to0$ the time evolution of a coherent state, which is
initially concentrated near a given point in phase space, is well
approximated by another coherent state, concentrated at the
classically evolved point. This statement is essentially what one
gets in the version of the present approach based on norm
convergences of states \cite{CLN} rather than norm convergence of
observables. What is missed in this approach are therefore the
operator properties (e) and (f).

\item{(F)}
{\it Limit of partition functions.}
\cite{Liebspin,Simon,Liebcoh,Wresin}
This aspect of the classical limit is conceptually straightforward,
because it only requires the convergence of some numbers. Of course,
it covers only a small fraction of the desirable features listed above.
Nevertheless some of the techniques developed for this problem, like
upper and lower symbols, or certain operators connecting spin
systems of different spin \cite{Liebcoh} are close to the approach
of this paper.

\item{(G)}
{\it Deformation quantization.}
\cite{Rieffel,RieffAMS,Landsman}.
In this approach the emphasis is indeed on the structure of products
and Poisson brackets, and it is in many ways close to ours. With
each classical phase space function (typically the Fourier transform
of a finite measure) one associates a specific family of
$\hb$-dependent operators, belonging to an algebra in which the
product is defined by some variant of the $\hb$-dependent Moyal
formula. It is clear that such families are also convergent
in our sense (see \Sec4.3). Nevertheless, the very restricted
$\hb$-dependence of such families is unnatural from the point of
view of the classical limit (or ``dequantization'' \cite{Emch}),
natural as it may be for ``quantization''. For another approach to
quantization, based on a very restricted class of Hamiltonians, see
\cite{Belliss}.

\bgsection 2. Definition and Main Results

Consider a typical Hamiltonian operator
$$ H\h= -{\hbar^2\over2m}\Delta +V(x)
\deqno(HV)$$
from a textbook on quantum mechanics. Our aim is to define the limit
of operators like $H\h$ as $\hb\to0$. Since the na\"\i ve approach
of setting $\hb=0$ in the above expression is obviously not what is
intended, we have to be more careful with the definition of such
limits. Rather than the algebraic expression \eq(HV), it must be
the relation of $H\h$ to other observables in the theory which has
to be taken to the limit. So let us denote by $\A\h$ the algebra of
observables ``at some value of $\hb>0$''. This will always be the
set of bounded operators on a Hilbert space (or a suitable
subalgebra), and hence in some sense independent of $\hb$. However,
the notational distinction between these algebras may help keeping
track of the various objects. Note that we will always consider
bounded observables. Thus it is not the operator \eq(HV) we will
take to the limit but, for example, its resolvent $(H\h-z)^{-1}$ or
the time evolution it generates.

For an $\hbar$-dependent observable $A\h\in\A\h$ we now want to define
``$\lim_{\hb\to0}A\h$''. Of course, since we have not yet put any
constraint on the allowed $\hb$-dependence of $A\h$, this limit
(whatever its definition) may fail to exist. The crucial notion we
must define is therefore ``$A\h$ converges as $\hb\to0$''. Loosely
speaking we must express the property that, for $\hb$ and $\hb'$ small
enough, $A\h$ and $A_{\hb'}$ become ``similar''. This shifts the
problem to the definition of some connection between the spaces $\A\h$
and $\A_{\hb'}$ which would permit such a comparison. The basic idea
of our approach is to use certain linear maps
$$ \jhh:\A_{\hb'}\to\A\h
\quad,\deqno(jhh)$$
and then to compare elements in the norm of $\A\h$.
Once the operators $\jhh$ are defined there will be no more
arbitrariness in the definition of the classical limit.

In order to illustrate this point, and to give a quick insight into
the kind of limits we will describe, we will proceed as follows:
{\it in this section we will assume that the spaces $\A\h$, and the
maps $\jhh$ have been defined}. Our aim is to show how this suffices
to set up a language, in which we can describe a limit with the
desirable features listed in the introduction. In particular, we
will state the main theorems of our approach in this subsection.
The actual definition of $\jhh$ will be given later, in the next
section, after the necessary preliminaries on phase space quantum
mechanics have been provided. In \Sec4 we will then be able to give
examples of convergent sequences of operators and states, by which
the reader will be able to judge whether we have indeed found a
rigorous statement of the usual folklore and intuitions on the
classical limit. Most proofs will be given in \Sec5, but those
relating to the dynamics had to be relegated to a sequel paper
\cite{CLD}.

The central notion of this paper is the following notion of
convergence, which we can define in terms of $\jhh$.

\iproclaim/D.conv/ Definition.
By an \mfat{\hbs} we mean a family of observables
$A\h\in\A\h$, defined for all sufficiently small $\hb$. We say that
an \hbs\ $A\h$ is \mfat{\jconv}, if
$$ \lim_{\hb'\to0}\limsup_{\hb\to0}
    \norm{A\h-\jhh A_{\hb'}}=0
\quad.$$
The set of \jconv\ \hbs\ will be denoted by $\CAj$.
Two \hbs s $A\h$ and $B\h$ are said to have the same limit, if
$$ \lim_{\hb\to0}\norm{A\h-B\h}=0
\quad.$$
Thus the {\bf limit} of $A\h$ is defined as an equivalence class of
\jconv\ \hbs s, and we will denote it by $\jlim\h A\h$, or
sometimes just $A_0$.
The space of all limits of \jconv\ \hbs s will be denoted by
$\A_0$.
\eproclaim

The abstract definition of $\jlim\h A\h$ as an equivalence class is
the best we can do without giving a concrete definition of $\jhh$. It
will be evident from our definition of $\jhh$, however, that the
limits can be identified with functions on phase space (see
\Def6/D.jhh/ and \Prp7/P.A0/). The convergence of operator products
to products of functions can then be stated as follows:

\iproclaim/T.prod/ Product Theorem.
Let $A\h, B\h$ be \jconv\ \hbs s, and define, for each $\hb$,
$C\h=A\h B\h\in\A\h$. Then $C$ is \jconv, and
$$ \jlim\h (A\h B\h)=(\jlim\h A\h)(\jlim\h B\h)
\quad,$$
where the product on the right hand side is the product in the
commutative algebra $\A_0$.
\eproclaim

Since the product in $\A_0$ is abelian, commutators
$\bracks{A\h,\B\h}$ are \jconv\ to zero. The interesting term for
commutators is thus the next order in $\hb$. It is clear, however,
that $\hb^{-1}\bracks{A\h,\B\h}$ cannot be \jconv\ for arbitrary
\jconv\ $A\h$ and $B\h$: any sequences $A\h,B\h$ with norm going to
zero are \jt-convergent, but this does not even suffice to force the
scaled commutators to stay bounded. Hence we need better control of
the \hbs s than mere \jt-convergence. A hint of the kind of
condition needed here is given by the theorem below: the Poisson
bracket to which these commutators converge is only defined for
differentiable limit functions. Hence we need differentiability
properties also for the sequences $A\h$ and $B\h$. The appropriate
space of sequences, denoted by $\C2(\A,\jt)$, will be defined and
discussed in \cite{CLD}. Briefly, $\C2(\A,\jt)$ consists of those
sequences $A\h$ such that $\pstr_{\epsilon\xi}(A\h)$ has Taylor
expansions to second order in $\epsilon$ with derivatives in $\CAj$
and an error estimate which is uniform for sufficiently small
$\hbar$. This space is norm dense in $\CAj$. The following theorem
is also shown in \cite{CLD}.

\iproclaim/T.poiss/ Bracket Theorem.
Let $A,B\in\C2(\A,\jt)$. Then $\hb^{-1}\bracks{A\h,B\h}$ is \jconv,
and
$$ \jlim\h{i\over\hb}\bracks{A\h,B\h}
      =\Pbrack{\jlim\h A\h,\ \jlim\h B\h}
\quad,$$
where the product on the right hand side is the Poisson bracket of
$\C2$-functions on phase space.
\eproclaim

Commutators and Poisson brackets determine the equations of motion
for quantum and classical systems, respectively. Hence the above
theorem says that the quantum equations of motion converge to the
classical ones. Of course, one also wants to know that the {\it
solutions} of the respective equations converge. This is the content
of the following Theorem. Again the proof is given in \cite{CLD}.
Note that the Theorem only makes a statement for finite times, \ie
it is not strong enough to allow the interchange the limits
$\hb\to0$, and the ergodic time average, or some other version of
the limit $t\to\infty$. This would be very interesting for
applications to ``quantum chaos'' (see \cite{Esposti} for a result
in this direction).

\iproclaim/T.dyn/ Evolution Theorem.
Let $H\h\in\C2(\A,\jt)$ such that $H\h=H\h^*$ for every $\hb$.
Define the time evolution for each $\hb$ by
$$ \tev t\hb(A)=e^{\textstyle itH\h /\hb}A\,
                e^{\textstyle-itH\h /\hb}
\quad,\deqno(tev)$$
for $A\in\A\h$, and $t\in\Rl$. Let $A\h$ be \jconv,
and define $A\h^t=\tev t\hb(A\h)$, for every $\hb$. Then $A^t\h$ is
also \jconv, and
$$ \jlim\h \tev t\hb(A\h) = \tev t0\bigl(\jlim\h A\h\bigr)
\quad,$$
where $\tev t0$ is the Hamiltonian time evolution on phase space
generated by the Hamiltonian function $H_0=\jlim\h H\h$.
\eproclaim

Finally, we would like to define the convergence of states. The
states for each $\hb$ are, by definition, positive, normalized
linear functionals on $\A\h$. Since $\A\h$ is an algebra of
operators on Hilbert space this includes all states given by density
matrices, the so-called normal states. Non-normal states appear
naturally in the description of limiting situations such as states
with sharp position and infinite momentum. They are also included in
the present setup.

\iproclaim/D.state/ Definition.
For each $\hb$, let $\omega\h:\A\h\to\Cx$ be a state. We say that
the \hbs\ $\omega$ is \mfat{\jsconv}, if for every \jconv\
\hbs\ $A\h\in\A\h$ of observables, the sequence of numbers
$\omega\h(A\h)$ has a limit as $\hb\to0$. The {\bf limit} of the
sequence is the state $\omega_0=\jslim\h\omega\h:\A_0\to\Cx$,
defined by
$$ \omega_0\bigl(\jlim\h A\h\bigr)
    =\lim_{\hb\to0} \omega\h(A\h)
\quad.$$
$\omega_0$ will be called a {\bf cluster point} of the sequence
$\omega\h$, if there is a subsequence $\hb_n, n\in N$ such that the
above equation holds for limits along this subsequence.
\eproclaim

Since $A_0=\jlim\h A\h$ is a function on phase space, the limit
functionals $\omega_0$ are measures on phase space, or, more
precisely, measures on a compactification of phase space. We will
see that every state on $\A_0$ occurs as the limit of suitable \hbs
s of states. \Def/D.state/ gives the analogue of weak*-convergence of
states on a fixed algebra. In particular, every sequence $\omega\h$
has cluster points. Norm limits of states will be considered in
another paper \cite{CLN}.

\bgsection 3. Definition of \bigfatjhh

Without the concrete definition of $\A\h$ and $\jhh$ the statements
made in the last section are void. In this section we will provide
these definitions, and describe some further properties of the limits,
which can be stated only in this more concrete context.

The systems we treat will be non-relativistic with $d<\infty$
degrees of freedom. Let us denote by $X=\Rl^d$ the configuration
space of the system. Then its {\it Hilbert space} is
$$ \H=\L2(X,dx)
\deqno()$$
In $\H$ we have a representation of the translations in
configuration space and momentum space, given by the unitary
{\it Weyl operators }
$$ \Bigl(\Weylg(x,p)\psi\Bigr)(y)
          = \exp\left(-{i\over2\hb}p\cdot x
                      +{i\over\hb} p\cdot y\right)\
               \psi(y-x)
\quad.\deqno(weyl)$$
This is a translation by the momentum $p\in\Rl^d$ and the position
$x\in\Rl^d$. Taken together these two determine a point in {\it
phase space} $\Ps$, usually denoted by $\xi=(x,p)$.
The basic commutation relations for the Weyl operators then read
$$\eqalignno{
  \Weylg(\xi)\Weylg(\eta)
     &=e^{\textstyle{i\over2\hb} \syf\xi\eta}\Weylg(\xi+\eta)
\quad,&\deqal(weylrel)\cr
\hbox{where}\qquad
   \syf x{p\,;x',p'}&=p\cdot x'-p'\cdot x
&\deqal(syf)}$$
is the usual symplectic form on phase space. The {\it phase space
translations} act on quantum observables, represented by  bounded
operators $A\in\B(\H)$, (resp.\ classical observables, represented by
bounded measurable functions $f\in\L\infty(\Ps)$) via
$$\eqalign{
      \pstr_\xi(A)\quad&=\Weylg(\xi)\,A\,\Weylg(-\xi)   \cr
     \hbzero{\pstr_\xi}(f)(\eta)&= f(\eta-\xi)
\quad.\cr}\deqno()$$
In either case, \ie for $\hb\geq0$, we get
$\pstr_{\xi+\eta}=\pstr_\xi\pstr_\eta$. The Weyl operators are
eigenvectors of the translations, \ie
$$ \pstr_\xi\Bigl(\Weylg(\eta)\Bigr)
      = e^{\textstyle {i\over\hb} \syf\xi\eta}\Weylg(\eta)
\quad.\deqno(weyltrans)$$

The {\it comparison maps} $\jhh:\B(\H)\to\B(\H)$ will be taken to be
positive in the sense that $A\geq0\Longrightarrow\jhh(A)\geq0$, and
unital, \ie $\jhh(\idty)=\idty$. These properties are simply
required by the statistical interpretation of quantum mechanics. The
essential condition is the one linking the comparison to the phase
space structure: we will demand that
$$ \jhh\circ\oprime{\pstr_\xi}=\pstr_\xi\circ\jhh
\quad.\deqno(itwine)$$
Note that the set of operators $\jhh$ satisfying these conditions
for fixed $\hb,\hb'$ is convex and, with any operator $\jhh$, also
contains the operator
$$ \tj_{\hb\hb'}=\int\rho(d\xi)\ \pstr_\xi\circ \jhh
\quad,$$
where $\rho$ is any probability measure on phase space. Obviously,
in order to get a sensible limit we must require that the origin of
phase space is not shifted around in some arbitrary way (so only
$\rho$ centered near the origin will be allowed in the above
formula), and that no large scale smearing out (with $\rho$ of very
large variance) is contained in $\jhh$. We won't go into making
these requirements precise in this paper (see, however, \cite{CLJ}).
The main point is that all systems of comparison maps satisfying
these requirements {\it define the same class of \jconv\ \hbs s} via
\Def/D.conv/. Since our whole theory is not based on the detailed
behaviour of $\jhh$, but only on the class of \jconv\ \hbs s, we are
free in this paper to make a somewhat arbitrary but explicit choice
of $\jhh$ for the sake of simple presentation. The equivalence to
other choices, including an essentially unique ``optimal'' one will
be shown in \cite{CLJ}. Our choice of comparison maps will have the
special property that it maps quantum to quantum observables (at
different value of $\hb$) via a classical intermediate step. It is
clear that something like this must be possible from the idea that
the comparison described by the $\jhh$ should be at least
asymptotically transitive.

Positive maps taking quantum observables to classical ones and
conversely are well-known \cite{Bopp,Simon,Davies,Takahashi,QHA}.
These maps depend on the choice of a normal state, which is usually
taken to be coherent, \ie the ground state of some harmonic
oscillator. Let
$$    \coh(x)=(\pi\hb)^{-d/4} \exp{-x^2\over 2\hb}
\deqno(coh)$$
be the ground state vector of the standard oscillator Hamiltonian
$$\Hosc={1\over2}\sum_i(P_i^2+Q_i^2)
\quad,\deqno(Hosc)$$
with $P_i=(\hb/i)\partial/\partial x_i$. By
$\Coh=\Ket\coh\Bra\coh$ we will denote the corresponding
one-dimensional projection.
Then we set, for $f\in\L\infty(\Ps)$, and $A\in\B(\H)$,
$$\eqalignno{\eqgroup(j0h0)
    \j0\hb(A)(x,p)
         &=\Bra\coh \Weylg(-x,-p)A\,\Weylg(x,p) \Ket\coh
&\deqal\lasteq.a(j0h)\cr
    \j\hb0(f) \qquad
        &=\int\psm xp  f(x,p)\
            \Weylg(x,p)\Ket\coh\Bra\coh\Weylg(-x,-p)
\quad.&\deqal\lasteq.b(jh0)\cr}$$
In terms of $\Coh$ we can write this as
$$\eqalignno{
    \j0\hb(A)(x,p)
         &=\tr\bigl(A\, \pstr_{x,p}(\Coh)\bigr)
&\deqal\lasteq.a'(J0h)\cr
    \j\hb0(f) \qquad
        &=\int\psm xp  f(x,p)\ \pstr_{x,p}(\Coh)
\quad.&\deqal\lasteq.b'(Jh0)\cr}$$
The integrals in \eq(jh0) or \eq(Jh0) are to be interpreted as weak
integrals, \ie we have to take matrix elements of the integral, and
compute it as a family of scalar integrals, which converge by virtue
of the ``square integrability of the Weyl operators'' (see
\cite{QHA}). One readily verifies that $\j0\hb$ and $\j\hb0$ both
take positive into positive elements, and preserve the respective
unit elements. Moreover, these maps transform the phase space
translations according to
$$ \j\hb0\circ \hbzero{\pstr_{\xi}}=\pstr_{\xi}\circ\j\hb0
\quad,\quad\hbox{and}\qquad
   \j0\hb\circ\pstr_{\xi}=\hbzero{\pstr_{\xi}}\circ\j0\hb
\quad.\deqno(itwine:0)$$

Since $\xi\mapsto\Weylg(\xi)$ is strongly continuous,
$\xi\mapsto\pstr_\xi(\Coh)$ is continuous in trace norm, which
implies that $\j0\hb A$ is a uniformly continuous function for any
$A\in\B(\H)$. Uniform continuity of a function $f$ can be expressed
as $\Norm\big{\hbzero{\pstr_\xi} f-f}\to0$ for $\xi\to0$, where we
have used the supremum norm of functions in $\L\infty(\Ps)$. The
same continuity argument applies to $\j\hb0$ and, indeed, all
operators of the form $A=\j\hb0f$ are uniformly continuous in the
sense that $\Norm\big{\pstr_\xi(A)-A}\to0$. With these preliminaries
we can now define $\jhh$, and also describe the ranges of these
maps.

\iproclaim/D.jhh/ Definition.
For $\hb,\hb'>0$, we set
$$    \jhh=\j\hb0\circ\j0{\hb'}:\B(\H)\to\B(\H)
\quad,\deqno(D.jhh)$$
where the maps $\j\hb0:\L\infty(\Ps)\to\B(\H)$ and
$\j0\hb:\B(\H)\to\L\infty(\Ps)$ are defined by equations \eq(Jh0)
and \eq(J0h). Together with the convention $\j00=\id$, the maps
$\jhh$ are thus defined for $\hb,\hb'\geq0$. From the above
discussion it follows that, unless $\hb=\hb'=0$, the range of $\jhh$
is contained in $\A\h$, where
$$\eqalignno{\eqgroup()
    \A\h     &= \Set\Big{A\in\B(\H)\ \stt
                 \lim_{\xi\to0}\norm{\pstr_\xi(A)-A}=0}
&\deqal\lasteq.a(D.Ah)\cr
     \A_0    &= \Set\Big{f\in\L\infty(\Ps)\stt
                 \lim_{\xi\to0}\norm{\hbzero{\pstr_\xi}(f)-f}=0}
\quad.&\deqal\lasteq.b(D.A0)}$$
\eproclaim

The space of observables ``at the value $\hb$'' (see the beginning
of \Sec2) can be taken as all of $\B(\H)$, independently of $\hb$.
However, since after one application of a comparison map $\jhh$ only
continuous elements play a role, we will usually take $\A\h$ from
\eq(D.Ah) as the space of observables. Note that this space is also
the same for all $\hb$. Other possible choices are briefly indicated
in \Sec4.3.

We have now used the symbol $\A_0$ for two different spaces, and we
have to justify this by showing that the space $\A_0$ of uniformly
continuous functions on $\Ps$ as defined in \eq(D.A0) is indeed a
concrete representation of the abstract limit space $\A_0$ appearing
in \Def/D.conv/. This will also justify our referring to the limits
$\jlim\h A\h$ as functions on phase space in the previous section.

\iproclaim/P.A0/ Proposition.
Let $A\h$ be a \jconv\ \hbs. Then $\j0\hb A\h$ is a norm
convergent sequence of functions in the space $\A_0$, as defined in
\Def/D.jhh/. The identification
$$ \jlim\h A\h \equiv \lim\h \j0\hb A\h $$
defines an isometric isomorphism between $\A_0$, and the abstract
limit space of \Def/D.conv/.
\eproclaim

It is suggestive at this point to try an alternative definition of
``convergence as $\hb\to0$'': the map $\j0\hb$ already takes operators
to functions, \ie quantum to classical observables, and the
convergence of these functions is at least implied by the definition
we have given. Hence we might try to take the uniform convergence
of  $\j0\hb A\h$ as a definition. We will see in \Sec4.5, however,
that with this definition the Product \Thm/T.prod/ would fail, so
with this restricted definition we would miss an important desirable
feature of the classical limit. The example in \Sec4.5 is an
operator which in a sense oscillates more and more rapidly as
$\hb\to0$. If we exclude this sort of oscillation by an
``equicontinuity'' condition, \ie if we make the uniform continuity
condition in $\A\h$ also uniform in $\hb$, the convergence of $\j0\hb
A\h$ indeed becomes equivalent to convergence in the sense of
\Def/D.conv/ (see\Thm8/T.conv/ below).

In order to state this precisely, we define the {\it modulus of
continuity} of $X\in\A\h$, $\hb\geq0$,   as the function
$\lambda\mapsto\mc\h(X,\lambda)$, with
$$  \mc\h(X,\lambda):=\sup\Set\Big{\norm{\pstr_{\xi}(X)-X}
                           \stt \xi^2\leq\lambda}
\quad,\deqno(mc)$$
where the ``square'' of a phase space translation $\xi=(x,p)$ is
defined by $\xi^2=x^2+p^2$. This involves some arbitrariness since
positions and momenta have different physical dimensions. Any choice
of the form $\lambda q^2+\lambda^{-1} p^2$
would have done just as well, except that the estimates involving
$\jhh$ look a bit simpler when the Euclidean norm``$\sqrt{\xi^2}$''
in phase space matches the oscillator Hamiltonian \eq(Hosc), whose
ground state $\coh$ enters the definition of $\jhh$.

Uniform continuity of $X\in\A\h$ is equivalent to
$\lim_{\lambda\to0}\mc\h(X,\lambda)=0$.
Moreover, the properties \eq(itwine) and \eq(itwine:0), together with
the norm estimate $\norm{\jhh X}\leq\norm{X}$ imply
$$ \mc\h(\jhh(X),\lambda)\leq \mc_{\hb'}(X,\lambda)
\qquad\hbox{for $\hb,\hb'\geq0$.}
\deqno(lipsh)$$
(Note that the cases $\hb=0$ and $\hb'=0$ are included).
Now, for a \jconv\ \hbs, $A\h$ is well approximated for
small $\hb$ by $\jhh(A_{\hb'})$, which has $\hb$-modulus of continuity
at most $\mc_{\hb'}(A_{\hb'},\lambda)$. This bound holds uniformly
for small $\hb$, thus excluding rapid oscillations of $A\h$ for small
$\hb$. This is the basic idea of the following characterization of
\jconv\ sequences. It will be our basic tool for verifying
\jt-convergence of the various sequences of observables in the
examples of the next section. It also gives a quantitative meaning
to the intuition that ``nearly classical'' observables are those
that change little on a classical phase space scale, \ie have small
modulus of continuity. Whenever all relevant observables in some
given physical situation satisfy this criterion, the classical limit
is a good approximation, and quantitative bounds of this type can
also be given, by following the proofs.
This intuition can also be used \cite{IHQ} to give a very direct
(although ``nonstandard'') definition of the classical limit, which
is essentially equivalent to the one given in this paper.

\iproclaim/T.conv/ Theorem.
A sequence of observables $A\h\in\A\h$ is \jconv, if and
only if the following two conditions hold:
\item{(a)}
$\j0\hb(A\h)\in\A_0$ converges uniformly as $\hb\to0$.
\item{(b)}
$A\h$ is {\bf equicontinuous} in the following sense:
for any $\epsilon>0$, we can find $\hb(\epsilon),\lambda(\epsilon)$
such that, for $\hb\leq\hb(\epsilon)$, and
$\lambda\leq\lambda(\epsilon)$, we have
$\mc_\hb(A\h,\lambda)\leq\epsilon$.
\eproclaim

The idea of introducing the maps $\jhh$ was to get a precise meaning
of ``$A\h$ and $A_{\hb'}$ are similar''. Of course, this relation
should be approximately transitive. This is expressed by the
following estimate. Its concrete form depends on the choice of the
coherent state \eq(coh) in the definition \eq(j0h0), and on \eq(mc).
Note that each of the three parameters $\hb$ in the theorem may take
the value zero.

\iproclaim/estim/ Theorem.
Let $\hb,\hb',\hb''\geq0$, and let $X\in\B(\H)$. Then
$$\eqalignno{
  \norm{(\j{\hb}{\hb''}-\jj{\hb}{\hb'}{\hb''})X}
    &\leq\int_0^\infty\!\!\!\mud(d\theta)\
               \mc_{\hb''}(X,2\hb'\theta)
&\deqal(estim)\cr
  \norm{X-\jj\hb0\hb X}
    &\leq\int_0^\infty\!\!\!\mud(d\theta)\
               \mc\h(X,2\hb\theta)
\quad, &\deqal(estim:0)\cr
\hbox{where}\qquad
  \mud(d\theta)&= {\theta^{d-1}\over(d-1)!}\
                  e^{\textstyle-\theta}\ d\theta
\quad.}$$
In particular, if $X\in\A_{\hb''}$, the norm \eq(estim) goes to zero as
$\hb'\to0$, uniformly in $\hb$.
\eproclaim

An important Corollary of \Thm/estim/ is the following construction
of \jconv\ sequences and \jsconv\ states. The sequences described in
(1) are called ``basic sequences''  in the theory of
``generalized inductive limits'' \cite{KAC,FGH,LMD}. Their
convergence is equivalent to the asymptotic transitivity
$\j\hb{\hb''}\approx\jj\hb{\hb'}{\hb''}$ of the comparison.

\iproclaim/C.basic/ Corollary.
\item{(1)}
Fix $\hb'\geq0$ and $X\in\A_{\hb'}$. Then $X\h=\jhh X$ is \jconv,
and
$$ \jlim\h \jhh X= \j0{\hb'}X
\quad.$$
\item{(2)}
Let $\omega:\A_0\to\Cx$ be a state, and define, for every $\hb>0$ a
state $\omega\h:\A\h\to\Cx$  by $\omega\h(X)=\omega(\j0\hb(X))$.
Then $\omega\h$ is \jsconv, and $\jslim\h\omega\h=\omega$.
\item{(3)}
An \hbs\ $\omega\h$ of states on $\A\h$ is \jsconv\ if and only if
the sequence $\omega\h\circ\j\hb0$ is weak*-convergent in the state
space of $\A_0$.
\eproclaim

Usually we are interested in normal states on $\A\h$, \ie states of
the form $\omega\h(A)=\tr D\h A$, where $D\h$ is a density matrix.
This excludes, for example, states with sharp position and infinite
momentum. (These can be obtained as the Hahn-Banach extensions of a
pure state on the algebra of uniformly continuous functions of
position alone, and assign zero probability to any finite momentum
interval). Similarly, on the classical side we often consider
states of the form $\omega_0(f)=\int\mu(d\xi)f(\xi)$, where $\mu$ is a
probability measure on phase space. Note that this is a strong
assumption on the state: there are many states on $\A_0$ which live
``at infinity'', \ie on the compactification points \cite{PHU} of the
spectrum space of $\A_0$. However, for those states for which
position and momentum are both finite with probability $1$, we get
the following somewhat simplified criterion for convergence. It is
analogous to the convergence theorems for characteristic functions
in probability theory (see, \eg \cite{Chung}). Recall that
$\C{}_0(\Ps)$ denotes the complex valued functions on $\Ps$
vanishing at infinity.

\iproclaim/P.normalc/ Proposition.
Let $\omega\h$ be an \hbs\ of normal states. Then the following
conditions are equivalent:
\item{(1)}
$\jslim\h\omega\h=\omega_0$ exists, and is a measure on phase space.
\item{(2)}
For every $f\in\C{}_0(\Ps)$, the limit
$\lim\h\omega\h(\j\hb0 f)=\omega_0(f)$ exists, and $\omega_0$ is
normalized, \ie
$\sup\set{\omega_0(f)\stt f\in\C{}_0(\Ps),\ f\leq1}=1$.
\item{(3)}
For all $\xi\in\Xi$, the limit
$\lim\h\omega\h(\Weylg(\hb\xi))=\widehat\omega_0(\xi)$ exists,
and $\xi\mapsto\widehat\omega_0(\xi)$ is a continuous
function.
\eproclaim

\bgsection 4. Examples and Miscellaneous Results

\Examp 1. Functions of position or momentum:
Let $f:\Rl^d\to\Rl$ be bounded and uniformly continuous, and let $F\h$
be the multiplication operator $(F\h\psi)(x)=f(x)\psi(x)$. Then $F\h$
satisfies the equicontinuity condition in \Thm/T.conv/. Moreover,
$\j0\hb(F)$ is the convolution of $f$ with a Gaussian of variance
proportional to $\sqrt\hb$. Hence, by the uniform continuity of $f$,
$$\bigl(\jlim\h F\h\bigr)(x,p)=f(x)
\quad.\deqno()$$
Similarly, let $\widetilde F\h=f(P)$, where $f$ is evaluated in the
functional calculus of the $d$ commuting self-adjoint operators
$P_k={\hb\over i}\,{\partial\over\partial x_k}$. (This is the same as
taking the Fourier transform, multiplying with $f(p)$, and
transforming back). Then
$$\bigl(\jlim\h \widetilde F\h\bigr)(x,p)=f(p)
\quad.\deqno()$$

\Examp 2. Weyl operators:
The Weyl operators \eq(weyl) play a fundamental role. They oscillate
too rapidly to be convergent (see \Sec4.5), but with a suitable
rescaling of the arguments they do converge. For fixed
$\hx,\hp\in\Rl^d$, we set
$$ \Weylo(\hx,\hp)
    = \Weylg(\hb \hx,\hb \hp)
    = e^{\textstyle-{i\hb\over2}\hx\cdot\hp}\quad
      e^{\textstyle i\hp\cdot Q}\
      e^{\textstyle-i\hx\cdot P}
\quad.\deqno(Weylh)$$
By the Product Theorem and the previous example, this converges
to the phase space function $\hbzero\Weylo(\hx,\hp)$, defined as
$$\eqalign{
    \hbzero\Weylo(\hx,\hp)(x,p)
       &=\exp\bigl(i(\hp\cdot x- \hx\cdot p)\bigr)
        =e^{i\syf\hx{\hp;x,p}}
\quad,\cr\hbox{or}\qquad
    \hbzero\Weylo(\eta)(\xi)
        &=e^{i\syf\eta\xi}
\quad.}\deqno(Weyl0)$$
The notational distinction between the two sets of Weyl operators
reflects a difference in interpretation: while the basic Weyl
operators $\Weylg(\xi)$ implement a symmetry transformation,
expectations of $\Weylo(\xi)$ determine the probability distribution
of position and momentum observables. This is precisely analogous to
the dual role of selfadjoint operators in \QM\ as generators of
one-parameter groups on the one hand, and as observables on the
other. These also differ by a factor $\hb$, \eg the generator of the
time evolution is not the observable $H$, but $H/\hb$. Of course,
this distinction is usually irrelevant ($\hb=1$!), but is crucial
in the classical limit (see also \Sec4.5 below).

\Examp 3. Integrals of Weyl operators:
Let $\mu$ be a finite (possibly signed) measure on $\Rl^{2d}$, and
define
$$ F\h(\mu)=\int\mu(d\eta) \Weylo(\eta)
\quad.\deqno()$$
By the previous example this is an integral of \jconv\
sequences with $\hb$-independent weights.
It is easy to check using the Dominated Convergence Theorem that
such sequences are also \jconv. Moreover, the limit is the integral
of the limits. In the present case we get the Fourier transform of
the measure $\mu$ (with a symplectic twist, because $\Ps$ and its
dual vector space are identified via $\syfdots$):
$$ F_0(\mu)(\xi)
     =\bigl(\jlim\h(F\h(\mu)\bigr)(\xi)
     =\int\mu(d\eta)\ e^{i\syf\eta\xi}
\quad.\deqno()$$
There are two interesting special cases: If $\mu$ happens to be
absolutely continuous with respect to Lebesgue measure, the
``quantum'' Riemann-Lebesgue Lemma \cite{QHA} asserts that
$F\h(\mu)$ is a compact operator for all $\hb$, and $F_0(\mu)$ is a
continuous function vanishing at infinity. On the other hand, if
$\mu$ is a sum of point measures, $F\h(\mu)$ is an element of the
CCR-algebra, \ie the C*-algebra generated by the Weyl operators,
and the limit function $F_0(\mu)$ is almost periodic. These
correspondences are a special case of a correspondence theorem
\cite{QHA,PHU} for general phase space translation invariant spaces
of operators and functions, respectively. This general result can be
used to set up limit theorems for a variety of subspaces of $\A\h$.

The sequences $F\h(\mu)$ with absolutely continuous $\mu$ of compact
support have been made the basis of a discussion of the classical
limit by Emch \cite{Emch,Emchbook}. In his approach each classical
observable $F_0$ thus has a unique \hbs\ of quantum observables
$F\h$ associated with it, which is also typical for ``deformation
quantization'' approaches \cite{Rieffel,RieffAMS,Rieffprep}. In our
approach this constraint becomes unnecessary, both from a technical
and from a conceptual point of view. Emch's main emphasis is on
defining the (weak) convergence of states with respect to this
particular set of sequences. The intersection between his
``classical states'', and our \jsconv\ states is described precisely
by \Prp/P.normalc/.

\Examp 4. Resolvents of unbounded operators:
By definition, \jconv\ sequences are uniformly bounded in norm,
which excludes the treatment of all standard quantum mechanical
Hamiltonians. As a substitute, however, we can consider the
resolvents of such operators. The following Theorem summarizes a few
basic facts of this approach to unbounded operators.

\iproclaim/T.res/ Theorem.
Let $H\h$ be an \hbs\ of (possibly unbounded) self-adjoint operators.
We call $H\h$ \mfat{\jconv\ in resolvent sense}, if
$R\h(z)=(H\h-z)^{-1}$ is \jconv\ for some $z\in\Cx$ with $\Im z\neq0$.
Then
\item{(1)}
$R\h(z)$ is \jconv\ for all $z$ with $\Im z\neq0$.
\item{(2)}
If $V\h$ is a \jconv\ sequence with $V\h=V\h^*$, and $H\h$ is
\jconv\ in resolvent sense, then $H\h+V\h$ is \jconv\ in resolvent
sense.
\eproclaim

\proof:
(1) By the resolvent equation we have
$$ R\h(z')=\sum_{n=0}^\infty (z'-z)^n\,R\h(z)^{n+1}
\quad,$$
provided that $\norm{(z'-z)R\h(z)}<1$, which by self-adjointness of
$H\h$ is guaranteed by $\abs{z'-z}<\abs{\Im z}$. Each term in this
sum is \jconv\ by the Product Theorem, and convergence is uniform
in $\hb$. This suffices to establish \jt-convergence of the sum.
Iterating this argument, we find \jt-convergence of $R\h(z)$ for all
$z'$ in the same half plane as the originally given $z$. Since $H\h$
is assumed to be self-adjoint, we also get \jt-convergence
of $R\h(\Bar z)=R\h(z)^*$.

(2) We can argue exactly as in (1), using the series
$$ (H\h+V\h-z)^{-1}= (H\h-z)^{-1}\sum_{k=0}^\infty
       \bigl(V\h(H\h-z)^{-1}\bigr)^k
\quad,$$
which converges uniformly in $\hb$, provided
$\norm{V\h}\abs{\Im z}^{-1}\leq\epsilon<1$ for small $\hb$.
This will be the case if $\abs{\Im z}>\norm{V_0}$. For other values
of $z$ the convergence follows by (1).
\QED

An immediate application is to Schr\"odinger operators:
the kinetic energy $H\h=-\hb^2/(2m)\Delta$ is \jconv\ in resolvent
sense by \Sec4.1, and if $V$ is a fixed uniformly continuous
bounded potential, we conclude, for $\Im z\neq0$:
$$\eqalign{
  \jlim\h
     &\left({-\hb^2\over 2m}\ \Delta +V(x) -z\idty\right)^{-1}
       =R_0(z) \cr
\hbox{with}\qquad
\bigl(R_0(z)\bigr)(x,p)
      &=\bigl({p^2\over 2m}+V(x) -z\idty\bigr)^{-1}
\quad.\cr}\deqno(resolv)$$

At first sight, it seems that the class of potentials for which this
result holds is much larger. Indeed, the same technique is used to
construct the Hamiltonian for {\it relatively} bounded
perturbations \cite{Kato}, \ie perturbations $V$ for which
$\norm{V(H-z)^{-1}}<1$ for large $z$.
The Coulomb potential is bounded relative to the Laplacian in this
sense. However, in the above application the Laplacian is scaled
down with a factor $\hb^2$, so this relative boundedness of $V$ with
respect to $H$ cannot be used {\it uniformly} in $\hb$, and this
destroys the proof.

It is easy to see that not only this particular method fails for the
attractive Coulomb potential, but the statement itself is false:
suppose that the potential $V$ is not bounded below, and let
$R(x,p)=(p^2+V(x)-z)^{-1}$ be the classical resolvent function at
$z\in\Cx$. If the resolvents of the corresponding Schr\"odinger
operators were \jconv, this function would have to be uniformly
continuous. This is impossible: Let $x_n$ be a sequence such that
$V(x_n)\to-\infty$, and let $p_n$ be a sequence such that
$p_n^2=-V(x_n)$. Then
$$ R(x_n,\, p_n+\epsilon)-R(x_n,p_n)
     =(2p_n\epsilon+\epsilon^2-z)^{-1}+z^{-1}
\quad.$$
For fixed $\epsilon$ the first term goes to zero, \ie
$\sup_{x,p}\abs{R(x,p+\epsilon)-R(x,p)}\geq \abs z^{-1}$, and hence
$R$ is not uniformly continuous. It should be noted, however, that
this negative result only concerns {\it norm} convergence. Singular
objects like the Coulomb resolvent may still be weakly convergent in
the sense dual to the norm convergence of states \cite{CLN}.

\Examp 5. Implementing unitaries never converge:
The time evolution, and all other symmetry transformations on
$\B(\H)$ are implemented by unitaries $U\h$ as $A\h\mapsto U\h A\h
U\h^*$. Suppose that $U\h$ is \jconv. Then we conclude with the
Product Theorem that
$\jlim\h U\h A\h U\h^*
    =(\jlim\h A\h)\abs{\jlim\h U\h}^2
    =\jlim\h A\h$.
In other words, the symmetry transformation becomes trivial in the
classical limit. On the other hand, the time evolution and many other
canonical transformations act non-trivially in the limit by the
Evolution \Thm/T.dyn/. Hence in all these cases the implementing
unitaries cannot converge.

An instructive special case is the phase space translation by
$\eta\neq0$. This clearly acts non-trivially in the limit, and is
implemented by $X\h=\Weylg(\eta)$.
We have
$$ \j0\hb(X\h)(\xi)=\exp\textstyle{i\over\hb}\syf\eta\xi
               \cdot\exp\textstyle{-1\over4\hb}\eta^2
\quad.\deqno()$$
This converges to zero, uniformly in $\xi$. Hence the criterion (a)
of \Thm/T.conv/ is satisfied, and would indicate the limit $X_0=0$.
But, of course, (b) is violated for this ``rapidly oscillating
operator'': we get
$$ \mc\h(X\h,\lambda)=\sup\Set\Big{\abs{e^{i\alpha}-1}
            \stt \alpha^2\leq \textstyle
                     {\lambda\over\hb^2}(\eta^2}
\quad.\deqno()$$
For fixed $\lambda\neq0$ this expression is equal to $2$ for all
sufficiently small $\hb$.
It is clear from this example, that a notion of convergence based on
\Thm/T.conv/.(a) alone would not satisfy the Product Theorem, and is
hence too weak for many applications (compare \Sec4.4, \Sec4.7, and
\Sec4.9).

\Examp 6. Point measures:
The operators $\Coh=\Ket\coh\Bra\coh$, which we have
used in the definition of $\j0\hb$ and $\j0\hb$ are not \jconv:
$(\j0\hb(\Coh)\bigr)(\xi)=\exp\bigl(-\xi^2/(2\hb)\bigr)$ converges
pointwise as $\hb\to0$, but not uniformly, (and not to a continuous
function). On the other hand, we can also interpret the operators
$\Coh$ as the density matrices of an \hbs\ of states $\omega\h$.
This sequence is \jsconv: for $A\in\CAj$ we have
$$ \lim\h\omega\h(A\h)
      =\lim\h\tr\bigl(\Coh A\h\bigr)
      =\lim\h \bigl(\j0\hb(A\h)\bigr)(0)
      = A_0(0)
\quad.\deqno(Coh2point)$$
Hence these states converge to the point measure at the origin.
More generally, we get from \Prp/P.normalc/ the following statement:
a sequence of normal states $\omega\h$ converges to the point measure
at the origin iff $\omega\h\bigl(\Weylo(\xi)\bigr)\to1$ for
every $\xi\in\Ps$.

In case each $\omega\h$ has finite second moments we can give a
simple and intuitive sufficient criterion for convergence to this
point measure. Consider the standard oscillator Hamiltonian $\Hosc$
\eq(Hosc). Then we claim the inequality
$$   {1\over2} \left(\Weylo(\xi)+\Weylo(\xi)^* \right)
     \geq \idty-\xi^2 \Hosc
\quad,\deqno(cosE)$$
interpreted as an inequality between quadratic forms. To prove this,
note that the inequality is unchanged under any symplectic linear
transformation leaving the metric $\xi^2$, and hence $\Hosc$
invariant. We may thus transform to a standard form in which only
one component, say the $p_1$-component of $\xi$ is non-zero. Then,
according to \eq(Weyl0), $\Weylo(\xi)=\exp(ip_1Q_1)$, and in the
functional calculus of $Q_1$, we find
$\Re\bigl(\Weylo(\xi)\bigr)
     =\cos(p_1Q_1)\geq \idty- p_1^2Q_1^2/2\geq\idty-\xi^2\Hosc$.
Evaluating now the inequality \eq(cosE) on a sequence of states, we
find that, if
$$ \omega\h(\Hosc)\longrightarrow 0
\quad\hbox{as}\quad\hb\to0 \quad,$$
then $\omega\h(\Weylo(\xi))\to1$  for all $\xi$, and hence
$\jslim\h\omega\h$ is the point measure at $0$ by the above
arguments.  It is shown in \cite{CLJ} that any such sequence
$\omega\h$ could have been used in the definition of $\jhh$ instead
of $\Coh$, without changing the class of convergent sequences.

\Examp 7. Eigenstates:
Let $H\h$ be a sequence of self-adjoint operators which are \jconv\ in
resolvent sense. Let $\lambda\h$ be a sequence of real numbers,
converging to $\lambda_0$, and let $\psi\h$ be an eigenvector with
$$ H\h\psi\h=\lambda\h\psi\h
\quad,\deqno()$$
for each $\hb>0$. Let $\omega\h(X)=\bra\psi\h,X\psi\h>$ be the
corresponding state on $\A\h$. Consider a cluster point $\omega_*$ of
this sequence of states, \ie the limit along a subsequence $\hb_n$.
Then by the Product Theorem we have
$$ \omega_*\Bigl(\abs{R_0(z)-(\lambda_0-z)^{-1}}^2\Bigr)
   =\lim_{\hb_n\to0}\omega_{\hb_n}
           \Bigl(\abs{R\h(z)-(\lambda\h-z)^{-1}}^2\Bigr)
   =0
\quad,$$
because $\omega_{\hb_n}$ is a sequence of eigenstates. It follows
that $\omega_*$, considered as a measure on (a compactification of )
phase space is supported by the level set
$$ \set{\xi\stt H_0(\xi)=\lambda_0}
\quad.$$

In the one-dimensional case, and when the dynamics associated with
$H\h$ is also \jconv, we can say more: then $\omega_0$ has to be
invariant under the phase flow generated by $H_0$. Hence it has to
be equal to the micro-canonical ensemble at energy $\lambda$ for the
classical Hamiltonian $H_0$. In particular, all cluster points of
$\omega\h$ coincide, and we have convergence.

\Examp 8. WKB states:
The basic states for the WKB-method are vectors of the form
$$ \phi\h(x)=\phi(x)e^{iS(x)/\hb}
\quad,\deqno(wkb)$$
with a fixed vector $\phi\in\L2(\Ps)$, and the ``action''
$S:\Rl^d\to\Rl$.
The distribution of ``position'' in these vectors is
$\abs{\phi(x)}^2$, independently of $\hb$, and the rapidly
oscillating phase determines the momentum. Asymptotic estimates of
expectation values in such states are traditionally evaluated using
the stationary phase method \cite{Maslov}. Since this typically
involves some partial integration, the technical conditions in such
results usually demand some smoothness of $\phi$ and $S$. In our
context we can get by with the minimal assumptions needed to even
state the asymptotic formula.

\iproclaim/WKB/ Theorem.
Let $\phi\in\L2(\Rl^d)$ with $\norm{\phi}=1$, and
let $S:\Rl^d\to\Rl$ almost everywhere differentiable.
Set $\omega\h(A)=\bra\phi\h,A\phi\h>$, with $\phi\h$ from \eq(wkb).
Then $\omega\h$ is \jsconv\ with limit $\omega_0$ given by
$$ \omega_0(f)=\int dx\ \abs{\phi(x)}^2\ f(x,\dS(x))
\quad.$$
\eproclaim

\proof:
The states $\omega\h$ are normal, and $\omega_0$ is a probability
measure on phase space. Hence we may apply \Prp/P.normalc/. In the
expression
$$ \omega\h(\Weylo(x,p))
     =\int dy\ \Bar{\phi(y)}\
      \exp i\Set\big{ {\hb\over2}x\cdot p +p\cdot y
                 -{1\over\hb}(S(y)-S(y-\hb x))}
                 \phi(y-\hb x)
$$
we may replace $\phi(y-\hb x)$ by $\phi(y)$: the error is bounded by
$\norm{\Weylg(\hb x,0)\phi-\phi}$, which goes to zero by strong
continuity of the translations on $\L2$. Since $\abs{\phi(y)}^2$ is
integrable, and independent of $\hb$, we may carry out the limit
under the integral by the Dominated Convergence Theorem. This gives
$$ \lim\h\omega\h(\Weylo(x,p))
     =\int dy\ \abs{\phi(y)}^2\
        \exp i\Set\big{ p\cdot y -x\cdot \dS(y)}
\quad.$$
The exponential can be written as $\hbzero\Weylo(x,p)(y,\dS(y))$,
which shows that
$\omega\h(\Weylo(x,p))\to\omega_0(\hbzero\Weylo(x,p))$ with the
$\omega_0$ given in the Theorem.
\QED

When $S$ is reasonably smooth, the set
$ \LagrM_S=\Set\Big{(x,\dS(x)) \stt x\in\Ps}$,
which contains the support of the measure $\omega_0$ is a {\it
Lagrange manifold} in phase space, \ie a manifold on which the
symplectic form vanishes. This property remains stable under time
evolution, whereas the uniqueness of the projection
$(x,\dS(x))\mapsto x$  from $\LagrM$ onto the configuration space is
obviously not stable. The points where this projection becomes
singular are called caustics, and play an important role in the time
dependent WKB-method \cite{Maslov}. At such points, and at the
turning points of a bound state problem, it may become more
profitable to play the same game with wave functions $\phi$ in
momentum representation, and a $p$-dependent action $S$. The limits
of such states can be treated exactly as above, so we will not do it
explicitly.

\Examp 9. Interference terms, and pure states
           converging to mixed states:
The WKB-states $\omega\h$ of the previous example are pure for every
non-zero $\hb$. Yet their limit is not a point measure, \ie the
limit is a mixed state. Is the funny support of the limit measure (the
Lagrange manifold) perhaps a consequence of this purity? Are the
limits of pure states always singular with respect to Lebesgue
measure, as \Sec4.7 also suggests? We will see in this
example that, to the contrary, any measure on phase space can be the
limit of a sequence of pure states.

The basic observation is that the classical limit annihilates certain
``interference terms''. The following Proposition describes a
general situation in which this happens. Recall that two states
$\omega,\omega'$ on a C*-algebra are called orthogonal, if
$\norm{\omega-\omega'}=2$, or, equivalently, if, for every
$\epsilon>0$, there is an element $0\leq F_\epsilon\leq\idty$ in the
algebra such that $\omega(F_\epsilon)\leq\epsilon$, and
$\omega'(F_\epsilon)\geq1-\epsilon$. On the abelian algebra $\A_0$
two states (measures) are orthogonal if they have disjoint supports,
but also if one is, say, a sum of point measures, and the other is
absolutely continuous with respect to Lebesgue measure on phase
space.

\iproclaim/P.interf/ Proposition.
Let $\phi\h,\psi\h$ be \hbs s of unit vectors such that the states
$\bra\phi\h,\cdot\phi\h>$ and $\bra\psi\h,\cdot\psi\h>$ are \jsconv\
with orthogonal limits. Then, for any \jconv\ \hbs\ $A\h$, we have
$$ \lim\h \bra\phi\h,A\h\psi\h>=0
\quad. $$
\eproclaim

\proof:
Let $\omega_\phi, \omega_\psi$ be the limit states of the sequences
in the Proposition. Pick $F_\epsilon$ such that
$\omega_\phi(F_\epsilon)\leq\epsilon$, and
$\omega_\psi(\idty-F_\epsilon)\leq\epsilon$, and let
$F_{\epsilon,\hb}$ be a \jconv\ \hbs\ with $\jlim\h
F_{\epsilon,\hb}=F_\epsilon$. Then
$$\eqalign{
  \abs{\bra\phi\h,A\h\psi\h>}
      &\leq \abs{\bra\phi\h,F_{\epsilon,\hb}A\h\psi\h>}
           +\abs{\bra\phi\h,A\h(\idty-F_{\epsilon,\hb})\psi\h>}
           +\abs{\bra\phi\h,\bracks{A\h,F_{\epsilon,\hb}}\psi\h>}
\cr   &\leq \norm{A\h} \norm{F_{\epsilon,\hb}\phi\h}
           +\norm{A\h} \norm{(\idty-F_{\epsilon,\hb})\psi\h}
           +\norm{\bracks{A\h,F_{\epsilon,\hb}}}
\quad.}$$
The first two terms converge to limits less than $\epsilon$ by the
choice of $F_\epsilon$, and the last term goes to zero by the
Product Theorem.
\QED

\iproclaim/C.pdense/ Theorem.
Let $\omega_0$ be a state on $\A_0$, represented by a probability
measure on phase space. Then there is an \hbs\ $\phi\h$ of unit
vectors such that $\omega_0$ is the limit of the \jsconv\ \hbs\
$\omega\h=\bra\phi\h,\cdot\phi\h>$ of pure states.
\eproclaim

\proof:
By \Prp/P.normalc/.(2) we have to construct $\phi\h$ such that
$$ \omega'\h(f)= \bra\phi\h,\j\hb0(f)\phi\h>
               \longrightarrow\omega_0(f)
\quad,\deqno*()$$
for all $f\in\C{}_0(\Ps)$.
Let $f_n\in\C{}_0(\Ps)$ be a norm dense sequence. Since the
states are uniformly bounded, it suffices to show
$\omega'\h(f_n)\to\omega_0(f_n)$ for all $n$. We will do this by
constructing a sequence of vectors $\phi\h$, and a sequence
$\hb(N)$, $N\in\Nl$, such that $\hb(N)\to0$ as $N\to\infty$, and
$$  \abs{\bra\phi\h,\j\hb0(f_n)\phi\h>
            -\omega_0(f_n)}\leq 2^{-N}
\quad,$$
for $n\leq N$, and $\hb\leq\hb(N)$. We first pick a state
$\dot\omega_N$, which is a sum of finitely many point measures
(supported on different points) such that
$\abs{\omega_0(f_n)-\dot\omega_N(f_n)}\leq 2^{-(N+1)}$.
We know from \Sec4.6 that we can find pure states converging to
any point measure, and combining these using \Prp/P.interf/ we find
vectors $\phi\h$ such that
$\abs{\bra\phi\h,\j\hb0(f_n)\phi\h> -\dot\omega_N(f_n)}
   \leq 2^{-(N+1)}$, for sufficiently small $\hb$.
These are the vectors that have the desired approximation property.
\QED

\Examp 10. Wigner functions:
The Wigner function \cite{Wigner}, or ``quasi-probability density''
of a state $\omega$ can be written as
$$ (\Wigner\omega)(\xi)
     = ({2/\hb})^d\ \omega\bigl(\pstr_\xi(\parity)\bigr)
\quad,\deqno(wigner)$$
where $(\parity\phi)(x)=\phi(-x)$ is the parity operator
\cite{Grossman}. Here we have chosen the normalization such that
formally, or with suitable regularization,
$(2\pi)^{-d}\int\!dx\,dp\ (\Wigner\omega)(x,p)=1$.
Of course, $\Wigner\omega$ is rarely positive \cite{Hudson,PWI}, and
in general not even integrable. Ignoring such technical quibbles,
however, as most of the literature on Wigner functions does, we get
a ``simplified'' formulation of the classical limit, and also an
interesting class of convergent states.

The modified definition of the classical limit is based on an
alternative definition of $\j0\hb$ and $\j\hb0$, namely as the usual
Wigner-Weyl quantization and dequantization maps. These can be
defined using the adjoint of \eq(wigner):
$$\eqalign{
 \int\nohb{\psm xp}\bigl(\Wigner\omega\bigr)(x,p)\
        \bigl(\jWigner_{0\hb}A\bigr)(x,p)
    &= \omega(A) \cr
  \jWigner_{\hb0}\qquad
    &=\left(\jWigner_{0\hb}\right)^{-1}\cr
  \jWigner_{\hb\hb'}\qquad
    &=\jWigner_{\hb0}\jWigner_{0\hb'}
\quad.}\deqno(jwigner)$$
{}From the observation that $\Wigner$ maps the state $\omega$ to the
measure on phase space with the same Fourier transform \cite{QHA}
(or Weyl transform) we get
$$ \jWigner_{\hb\hb'}\Bigl(\oprime\Weylo(\xi)\Bigr)
      =\Weylo(\xi)
\quad.\deqno(Fwigner)$$
Because $\Wigner\omega$ is in general not integrable, the
transformations \eq(jwigner) are all ill-defined as they stand, and
unbounded for the norms of $\B(\H)$ and $\L\infty(\Ps)$
\cite{Daubechies}. There are several ways to give them a meaning on
some restricted domain. For example, all transformations make sense
on the Hilbert-Schmidt class and $\L2(\Ps)$, because $\Wigner$ is
unitary up to a factor:
$$ \int{\nohb{\psm xp}}\Bar{\bigl(\Wigner\omega\bigr)(x,p)}
       \bigl(\Wigner\omega'\bigr)(x,p)
     = \hb^{-d}\tr\bigl(D_{\omega}^*D_{\omega'}\bigr)
\quad,\deqno(wignerU)$$
where $D_\omega$ and $D_{\omega'}$ are the density matrices of
$\omega$ and $\omega'$. Further customary domains of such
transformations involve additional smoothness assumptions
\cite{Robert}. Whether one wants to burden the definition of
the classical limit with such constraints is a matter of taste.
That they are not necessary is demonstrated by the present paper, or
so the author hopes.

The transformation $\Wigner$ can also be used directly to define a
sequence of states $\omega\h$ by fixing a density $\rho\in\L1(\Ps)$,
and demanding
$$ \Wigner\omega\h=\rho
\quad.\deqno(wignerstate)$$
We have to assume that $\rho$ is chosen so that  $\omega\h$ is given
by a trace class operator $D\h$. Let us denote the the classical
state  with density $\rho$ by $\omega_0$. Is this the classical
limit of the sequence $\omega\h$, \ie  do we have
$$ \jslim\h\omega\h=\omega_0
\quad?\deqno(wigslimit)$$
In order to decide this, let us first consider a \jconv\ sequence of
quantum observables of the special form $A\h=\j\hb0A_0$. Then
$$\eqalign{
  \omega\h(A\h)&=\tr(D\h A\h) \cr
               &=\int\psm xp A_0(x,p)\ \tr(D\h
                  \pstr_{x,p}(\Coh))    \cr
               &= \int\nohb{\psm xp}\nohb{\psm{x'}{p'}}
                    A_0(x,p)K\h(x-x',p-p') \rho(x',p')
\quad,\cr}$$
where we have evaluated the trace using \eq(wignerU), and have used
the Wigner function
\break
$K\h(x,p)=(2/\hb)^d\exp\bigl(-(x^2+p^2)/\hb\bigr)$
of the coherent projection $\Coh$. This kernel goes to a
$\delta$-function as $\hb\to0$, and since $A_0$ is uniformly
continuous, $\omega\h(A\h)$ converges to
$(2\pi)^{-d}\int\!dx\,dp\, A_0(x,p)\rho(x,p)=\omega_0(A_0)$. So it
appears that \eq(wigslimit) holds.

What makes this computation work is the fact that the convolution of
two Wigner functions of trace class operators (here $D\h$ and
$\Coh$) is always integrable. Thus the bad properties of $\omega\h$
are averaged out. The argument fails, however, when $A\h$
is not of the special form $A\h=\j\hb0A_0$. It is true that the
$A\h$ of this form are norm dense ($\norm{A\h-\j\hb0A_0}\to0$ for
any \jconv\ sequence). However, this kind of approximation for a
general $\A\h$ is only sufficient to show the convergence of
$\omega\h(A\h)$, when $D\h$ is uniformly bounded in trace norm as
$\hb\to0$. Only in this case the conclusion \eq(wigslimit) is valid.
It is easy to find densities $\rho$, however, such that $D\h$ is not
even trace class for any $\hb$ (any density $\rho$, which is
unbounded, or discontinuous, or does not go to zero at infinity will
do). For such densities the sequence $\omega\h(A\h)$ may diverge,
even if $\norm{A\h}\to0$. In particular the Ansatz \eq(wignerstate)
with such $\rho$ never yields a \jsconv\ sequence $\omega\h$.

If we value the statistical interpretation of \QM, we should demand
not only that $D\h$ has uniformly bounded trace norm, but also that
$D\h$ (and hence $\omega\h$) is positive for all $\hb$, or at least
for a sequence $\hb_n$ along which we want to take the classical
limit. In the terminology of Narcowich \cite{Narcowich} this means
that the ``Wigner spectrum'' of the Fourier transform of $\rho$
contains the sequence $\hb_n$. This is a severe constraint on the
classical densities $\rho$ \cite{PWI}.

\bgsection 5. Proofs

In this section we prove the results stated in \Sec2 and \Sec3,
apart from the Theorems \pcl/T.poiss/ and \pcl/T.dyn/ about Poisson
brackets and the dynamics \cite{CLD}. We first state a Lemma that
allows us to handle the maps $\jhh$ with $\hb,\hb'\geq0$ and their
compositions more easily. The basic observation is that since all
these maps are normal, they are completely determined by their
action on Weyl operators. Moreover, since Weyl operators are
eigenvectors of the phase space translations \eq(weyltrans), and
these are intertwined by $\jhh$, Weyl operators must be mapped into
Weyl operators--- up to a scalar factor. This scalar factor is what
distinguishes $\jWigner_{\hb\hb'}$ after \eq(Fwigner) from $\jhh$,
and some such a factor is necessary to make $\jhh$ positive (see
\cite{CLJ} for a complete discussion).

\iproclaim/L.Fjhh/ Lemma.
For $\hb\geq0$, let $\Weylo(x,p)$ be defined as in \eq(Weylh),
and let $\jhh$ be as defined in equations \eq(j0h0) and \eq(D.jhh).
Then, for $\hb,\hb'\geq0$, and $A\in\B(\H)$, we have
$$\eqalignno{
  \jhh\oprime\Weylo(x,p)
      &= \Weylo(x,p)\ \exp{{-(\hb+\hb')\over4}(x^2+p^2)}
\quad,&\deqal(Fjhh)\cr
\noalign{\noindent and}
   \jj\hb{\hb'}\hb(A)&=\int{dx \,dp \over(2\pi(\hb+\hb'))^d}\
                     \exp{-(x^2+p^2)\over2(\hb+\hb')}\
                     \pstr_{x,p}(A)
\quad,&\deqal(jjconvol)}$$
\eproclaim

\proof:
A Gaussian integration using \eq(coh) and \eq(weyl) gives
$$  \bra\coh,\Weylg(x,p)\coh>
      =\exp{-1\over4\hb}(x^2+p^2)
\quad.\deqno(cWc)$$
{}From this and the Weyl relations \eq(weylrel) we get the equation
\eq(Fjhh) for the special case $\hb=0$. The case $\hb'=0$ is
verified by the following computation:
$$\eqalign{
  \j\hb0\hbzero\Weylo(x,p)
      &= \int\psm {x'}{p'} \exp\,i(x'\cdot p-p'\cdot x)\
             \pstr_{x',p'}(\Coh)  \cr
       &= \Weylg(\hb x,\hb p)\ \int\psm {x'}{p'}
             \pstr_{x',p'}\Bigl(\Weylg(\hb x,\hb p)^*\Coh\Bigr) \cr
      &= \Weylo(x,p)\ \tr\bigl(\Weylg(\hb x,\hb p)^*\Coh\bigr)\cr
      &= \Weylo(x,p)\ \exp{-\hb\over4}(x^2+p^2)
\quad.}$$
For general $\hb,\hb'$ we get \eq(Fjhh) by composition. In the same
way we find
$\jj\hb{\hb'}\hb\Weylo(x,p)
    =\Weylo(x,p)\ \exp{-(\hb+\hb')\over2}(x^2+p^2)$.
When $A$ is a Weyl operator, \eq(jjconvol) can be verified by computing
the Gaussian integral. For other operators $A$, \eq(jjconvol)
follows, because $\jj\hb{\hb'}\hb$ is ultraweakly continuous, and
the Weyl operators span an irreducible algebra of operators, which
is hence ultraweakly dense in $\B(\H)$.
\QED

\proof{ of \Thm/estim/:}
{}From \Lem/L.Fjhh/ and the definition \eq(mc) of
the modulus of continuity we get
$$\eqalign{
   \norm{f-\jj0\hb0f}
      &=\sup_{x,p}\abs{\int \psm{x'}{p'}
                 \exp{-1\over2\hb}\Bigl((x-x')^2+(p-p')^2\Bigr)\
                 \bigl(f(x,p)-f(x',p')\bigr)} \cr
      &\leq\int \psm xp
                 \exp{-1\over2\hb}\Bigl(x^2+p^2\Bigr)\
                 \mc_0(f,(x^2+p^2))  \cr
      &=\int_0^\infty\!\!\!d\theta\,
                 {\theta^{d-1}\over(d-1)!} e^{-\theta}\
                  \mc_0(f,2\hb\theta)
\quad.\cr}$$
The estimate \eq(estim) now follows from
$$ \norm{(\j{\hb}{\hb''}-\jj{\hb}{\hb'}{\hb''})X}
    =\norm{\j\hb0(\id-\jj0{\hb'}0)\j0{\hb''}X}
\quad,$$
$\norm{\j\hb0X}\leq\norm{X}$, and
$\mc_0(\j0{\hb''}X,\lambda)\leq\mc_{\hb''}(X,\lambda)$.
The proof of the estimate \eq(estim:0) is completely analogous.
\QED

\proof{ of \Thm/T.conv/:} Assume that $A\h$ is \jconv.
We first show the equicontinuity condition (b). By \Def/D.conv/ we
can pick $\hb'$ such that \break
$\limsup\h\norm{A\h-\jhh A_{\hb'}}\leq\epsilon/8$.
Next we pick $\hb(\epsilon)$ such that
$\norm{A\h-\jhh A_{\hb'}}\leq\epsilon/4$ for $\hb\leq\hb(\epsilon)$.
Since $A_{\hb'}\in\A_{\hb'}$, we can find $\lambda(\epsilon)$ such
that $\mc_{\hb'}(A_{\hb'},\lambda)\leq\epsilon/2$ for
$\lambda\leq\lambda(\epsilon)$.
Hence, for $\hb\leq\hb(\epsilon)$, and $\lambda\leq\lambda(\epsilon)$,
$$\eqalign{
   \mc\h(A\h,\lambda)
      &\leq \mc\h(\jhh A_{\hb'},\lambda)
              +2\norm{A\h-\jhh A_{\hb'}} \cr
      &\leq \mc_{\hb'}(A_{\hb'},\lambda)+ \epsilon/2
       \leq\epsilon
\quad.}$$

To see condition (a), the uniform convergence of $\j0\hb A\h$, we
estimate
$$\eqalign{
   \norm{\j0\hb A\h- \j0{\hb'}A_{{\hb'}}}
     &\leq\norm{\j0\hb(A\h-\jhh A_{\hb'})}
         +\norm{(\jj0\hb{\hb'}-\j0{\hb'})A_{\hb'}} \cr
     &\leq\norm{A\h-\jhh A_{\hb'}}
         +\norm{(\id-\jj0\hb0)\j0{\hb'}A_{\hb'}}
\quad.}$$
Since $\j0{\hb'}A_{\hb'}$ is uniformly continuous, the second term
goes to zero as $\hb\to0$. Hence, using the \jt-convergence of $A\h$ for
the first term, we get
$\lim_{\hb'}\limsup\h\norm{\j0\hb A\h- \j0{\hb'}A_{{\hb'}}}=0$,
which implies that $\j0\hb A\h$ is norm-Cauchy in $\A_0$, and hence
converges.

Conversely, assume that (a) and (b) are satisfied. Then
$$\eqalign{
    \norm{A\h-\jhh A_{\hb'}}
     &\leq \norm{A\h-\jj\hb0\hb A\h}
         + \norm{\jj\hb0\hb A\h- \jj\hb0{\hb'} A_{\hb'}} \cr
     &\leq \int\mud(d\theta) \mc\h(A\h,2\hb\theta)
           \quad+\quad \norm{\j0\hb A\h- \j0{\hb'}A_{\hb'}}
\quad}$$
The integrand in the first term goes to zero as $\hb\to0$, for every
$\theta$ due to condition (b), so the first term vanishes in this
limit by dominated convergence.  Hence by condition (a),
$\lim_{\hb'}\limsup\h\norm{A\h- \jhh A_{\hb'}}=0$.
\QED

\proof{ of \Prp/P.A0/:}
Let us denote, for the sake of this proof, the
abstract limit space of \Def/D.conv/ by $\A_\infty$, and the space
of uniformly continuous functions from \eq(D.A0) by $\A_0$.
Then the equation
$$ \j0\infty\bigl(\jlim\h A\h\bigr)= \lim\h\j0\hb A\h
\quad,$$
for \jconv $A\h$, defines an operator
$\j0\infty:\A_\infty\to\A_0$, because $\jlim\h A\h=0$ is defined as
$\lim\h\norm{A\h}=0$, and hence implies $\lim\h\j0\hb A\h=0$.
$\j0\infty$ is surjective, because $\jlim\h\j\hb0f=f$ for
$f\in\A_0$, and is injective by the estimate \eq(estim:0). Since
both $\j0\hb$ and $\j\hb0$ are contractive the same arguments also
show that $\j0\infty$ is isometric.
\QED

\proof{ of \Cor/C.basic/:}
$\jt$-convergence in (1) follows immediately from the Theorem, and
the value of the limit follows from the identification of the limit
space. For (2) it suffices to evaluate $\omega\h$ on \jconv\
sequences of the form (1), for which the convergence again follows
from \eq(estim) with $\hb=0$. (3) follows from the observation that
$$ \lim\h\norm{A\h-\j\hb0A_0}=0
\quad,\deqno()$$
for any $A\in\CAj$.
\QED

\proof{of \Prp/P.normalc/:}
(1)$\Rightarrow$(3):
By \Sec4.2, $\Weylo(\xi)=\Weylg(\hb\xi)$ is
\jconv, hence the existence of the limit is clear, which is then equal
to $\omega_0(\hbzero\Weylo(\xi))$. This is the Fourier transform of the
measure $\omega_0$, which is continuous by Bochner's Theorem \cite{Boch}.

(3)$\Rightarrow$(2):
Positivity of $\omega\h$ is equivalent \cite{QHA,PWI} to the
positive definiteness of all matrices $M_{\nu\mu}^\hb$,
$\nu,\mu=1,\ldots,N$ defined by
$$ M_{\nu\mu}^\hb
        =\omega\h\bigl(\Weylo(\xi_\nu-\xi_\mu)\bigr)\
          e^{\textstyle i\hb\syf{\xi_\nu}{\xi_\mu}}
\quad,$$
for all choices of $\xi_1,\ldots,\xi_N\in\Ps$. In the limit
$\hb\to0$ this becomes the positive definiteness hypothesis in
Bochner's theorem, which together with the postulated
continuity implies that $\widehat\omega_0(\xi)$ is the Fourier
transform of a positive measure $\omega_0$ on $\Ps$, \ie
$\widehat\omega_0(\xi)=\omega_0\bigl(\hbzero{\Weylo(\xi)}\bigr)$.
The normalization of this measure follows by setting $\xi=0$.

It suffices to show convergence for $f=F_0$ in a norm dense subset of
$\C{}_0(\Ps)$. For this we take the Fourier transforms of
$\L1$-functions in the sense of \Sec4.3.
Explicitly, we let $f=\int\!d\xi\,\rho(\xi)\hbzero{\Weylo(\xi)}$
with fixed $\rho\in\L1(\Ps)$. Then
$$\eqalign{
     \j\hb0&=\int\!d\xi\ \rho(\xi)
                    e^{-\hb\xi^2/4}\ \Weylo(\xi)
\quad,\cr\hbox{and}\qquad
   \omega\h(\j\hb0f)
     &=\int\!d\xi\,\rho(\xi)e^{-\hb\xi^2/4}\ \omega\h
       \bigl(\Weylo(\xi)\bigr)
}$$
holds for all $\hb\geq0$, and the claim follows by dominated
convergence.

(2)$\Rightarrow$(1):
By \Cor/C.basic/.(3) we have to show that the convergence
$\omega\h(\j\hb0f)\equiv\omega'\h(f)\to\omega_0(f)$
extends from $f\in\C{}_0(\Ps)$ to all $f\in\A_0$.
By the normalization condition in (2), we can find
$f_\epsilon\in\C{}_0(\Ps)$ such that $0\leq f_\epsilon\leq1$, and
$\omega_0(f)\geq1-\epsilon$. Hence
$\omega'\h(f_\epsilon)\geq1-2\epsilon$ for $\hb\leq\hb(\epsilon)$.
But then, for arbitrary $f\in\A_0$,
$$  \abs{\omega'\h(f)-\omega_0(f)}
        \leq \abs{\omega'\h(f(1-f_\epsilon)}
             +\abs{\omega'\h(ff_\epsilon)-\omega_0(ff_\epsilon)}
             +\abs{\omega_0(f(1-f_\epsilon)}
\quad.$$
Then, for $\hb\leq\hb(\epsilon)$, the first and last term are
bounded by $2\epsilon\norm{f}$ and $\epsilon\norm{f}$, respectively,
and the middle term goes to zero, because
$ff_\epsilon\in\C{}_0(\Ps)$.
\QED

\proof{ of the Product \Thm/T.prod/:}
Let $A\h,B\h$ be convergent \hbs s. We have to show the convergence
of $C\h=A\h B\h$. Two observations help to simplify the proof: firstly
we may replace $A\h$ by $A\h'$ such that $\norm{A\h-A\h'}\to0$, and
similarly for $B\h$. Note that this modification will also not change
$\jlim\h C\h$. Hence we may take $A\h=\j\hb0A_0$, and $B\h=\j\hb0B_0$.
Secondly, the estimate
$$ \mc\h(A\h B\h,\lambda)
     \leq \mc\h(A\h,\lambda)\norm{B\h}+\norm{A\h}\mc\h(B\h,\lambda)
\quad$$
shows that $C\h$ satisfies the equicontinuity condition in
\Thm/T.conv/, since $A\h$ and $B\h$ do. Therefore, by  that
Theorem, it suffices to show that, for $A_0,B_0\in\A_0$,
$$ \lim\h\norm{\j0\hb\bigl((\j\hb0A)(\j\hb0B)\bigr) - A_0B_0}
      =0
\quad.$$
This norm is the supremum norm in the function algebra $\A_0$, hence
it suffices to estimate it at any point, say the origin,  in terms
of data, which do not change under translation. Specifically, we
will give a bound on
$$ \abs{\j0\hb\bigl((\j\hb0A_0)(\j\hb0B_0)\bigr)(0) - A_0(0)B_0(0)}
\quad\deqno*()$$
by a quantity depending only on moduli of continuity of $A_0$ and
$B_0$. Then \eq*() is bounded by
$$\eqalign{
   &\abs{\j0\hb\bigl(\j\hb0(A_0-A_0(0)\idty)
             \j\hb0(B_0-B_0(0)\idty)\bigr)(0)}    \cr
   &\qquad+\abs{A_0(0)}\norm{B_0-\jj0\hb0B_0}
          +\norm{A_0-\jj0\hb0A_0}\abs{(\jj0\hb0B_0)(0)}
\quad,}$$
where the terms in the second line go to zero by virtue of \eq(estim).
Hence in \eq*() we may suppose that $A_0(0)=B_0(0)=0$
and, consequently, $\abs{A_0(\xi)}\leq\mc_0(A_0,\xi^2)$ . Inserting the
definitions \eq(J0h) and \eq(Jh0) of $\j0\hb$ and $\j\hb0$, we obtain
$$\eqalign{
    &\abs{\j0\hb\bigl((\j\hb0A_0)(\j\hb0B_0)\bigr)(0)}    \cr
    &\qquad\leq \int \psm xp \psm{x'}{p'}
         \abs{A_0(x,p)}\,\abs{B_0(x',p')}\,
         \abs{\tr\bigl(\Coh\pstr_{x,p}(\Coh)
                           \pstr_{x',p'}(\Coh)\bigr) }    \cr
    &\qquad\leq \int \psm xp \psm{x'}{p'}
         \mc_0(A_0,x^2+p^2) \,
         \mc_0(B_0,x^{\prime2}+p^{\prime2})   \ \times
        \cr  &\hskip80pt   \times\
         \abs{\bra\coh,\Weylg(x,p)\coh>
              \bra\coh,\Weylg(x,p)^*\Weylg(x',p')\coh>
              \bra\coh,\Weylg(x',p')^*\coh>}              \cr
    &\qquad\leq \int \psm xp \psm{x'}{p'}
         \mc_0(A_0,x^2+p^2) \,
         \mc_0(B_0,x^{\prime2}+p^{\prime2})\,      \ \times
        \cr  &\hskip80pt   \times\
         \exp{-1\over4\hb}(x^2+p^2)\
         \exp{-1\over4\hb}(x^{\prime2}+p^{\prime2})      \cr
    &\qquad\leq 2^{2d}
         \int\mud(d\theta) \mc_0(A_0,4\hb\theta) \
         \int\mud(d\theta')\mc_0(B_0,4\hb\theta')
\quad.\cr}$$
Note that we are justified in using the weak* integrals defining
$\j\hb0$ because in both integrations in the second line the
definition of $\j\hb0$ is used under the trace with a trace class
operator. In any case, since the integrals in the last line go to
zero by dominated convergence, we find that
$\abs{\j0\hb\bigl((\j\hb0A_0)(\j\hb0B_0)\bigr)(0)}\to0$. The
estimate involves only the moduli of continuity of $A$  and $B$,
which concludes the proof.
\QED

\bgsection 6. Further extensions

\item{(1)}{\it Alternative definitions of $\jhh$}\hfill\break
It was claimed in \Sec2 that the precise definition of $\jhh$ is
not essential, since asymptotically close systems of comparison maps
yield the same class of \jconv\ \hbs s. In \cite{CLJ} the class of
alternative choices of $\jhh$ with this property is studied
systematically. Surprisingly, there is even one choice for which the
chain relation $\j\hb{\hb''}=\jj\hb{\hb'}{\hb''}$ is satisfied
exactly.  Hence the classical limit can be understood as an ordinary
inductive limit of ordered normed spaces. The ``sharpest possible''
comparison maps satisfying the chain relation are essentially unique
(\ie up to the choice of a complex structure on phase space).

\item{(2)} {\it Norm limits of states}\hfill\break
It is easy to define comparison maps $\tj_{\hb\hb'}$ for density
matrices, which determine the notion of a norm convergent sequence
of states in the classical limit: for example we may take
$\tj_{\hb\hb'}$ as the pre-adjoint of $\j{\hb'}\hb$ from \Sec3.
The limit space then consists of all integrable functions on phase
space \cite{CLN}. Also the evaluation of a norm convergent sequence
of states on a norm convergent sequence of observables produces a
convergent sequence of numbers, or, what is the same thing, norm
convergence of either states or observables implies weak
convergence.  The notion of weak convergence of observables in the
classical limit allows one to discuss, for example, the convergence
of spectral projections. However, the Product Theorem is lost for
this weak convergence. The limits of WKB states or eigenstates (see
\Sec4.7 and \Sec4.8) do {\it not} exist in norm, since the limit
measures are not absolutely continuous. On the other hand, under
suitable conditions the equilibrium states belonging to a norm
convergent sequence of Hamiltonians do converge in norm.

\item{(3)} {\it Dynamics }\hfill\break
The definition of the class $\C2(\A,\jt)$, as well as the proofs of
\Thm/T.poiss/ and \Thm/T.dyn/ will be given in \cite{CLD}. As
written, these theorems require bounded Hamiltonians, which comes
from the technical requirement that the time evolution should be
strongly continuous on $\A\h$. Not even the time evolution of the
free particle satisfies this. On the other hand, by restricting
$\A\h$ to the space of compact operators with adjoined identity,
certain unbounded Hamiltonians can be treated, as well. Note,
however, that a version of the Evolution Theorem can only hold if
the classical time evolution exists for all times, so some
restrictions on $H\h$ are always needed. A good way of handling
unbounded Hamiltonians is also to study the dynamics in the norm
limit of states (see (2), and \cite{Hepp,Hagedorn}). In the
deformation quantization approach, dynamics was recently discussed
in \cite{Rieffprep}.

\item{(4)} {\it Classical trajectories}\hfill\break
The Evolution Theorem does not explain how, in the classical limit,
a description of the systems in terms of trajectories becomes
possible. The statistics of trajectories should be the limit of a
sequence of continual measurement processes depending on $\hb$.
One can set up such processes quite easily in the framework of E.B.
Davies \cite{Davies}, and thus obtains an idealized description of a
measuring device which is always in interaction with the system
under consideration, and produces as output a sequence of random
events, each of which is described by a Poisson distributed random
time, and a random point in phase space. The rate $\nu$ of random
events can be chosen arbitrarily, but it is clear that a larger rate
will introduce a stronger perturbation of the free evolution. What
happens in the classical limit now depends on how this rate $\nu$ is
scaled as $\hb\to0$. If we take $\nu\to\infty$, but $\nu\hb\to0$, we
will get a classical process, which is concentrated on the classical
orbits, and has the initial condition as the only random parameter.
On the other hand, if we take $\nu\hb\to C$, some quantum
perturbations of the free evolution survive the limit, and we get a
diffusion in phase space with diffusion constant proportional to
$C$ (compare \cite{Hajo}), and with a drift given by the Hamiltonian
vector field. Joint work on these issues is in progress with Fabio
Benatti.

\item{(5)} {\it Higher orders in $\hb$}\hfill\break
In the WKB method one is usually not only interested in the classical
limit, but in the asymptotic expansion of the wave functions to all
orders in $\hb$. In this paper we have only considered the limit
itself, for the following reason: we wanted to emphasize that the
notion of convergence is almost completely insensitive to special
choices of identification operators $\j\hb{\hb'}$.
Given these identifications one can also define higher orders of the
asymptotic expansion  of an $\hb$-dependent operator.  But these are
now much less ``canonical'', and there seems little point in
computing such quantities which depend on a special choice of, say,
coherent states, unless there is a specific reason for considering a
particular choice. One possible ``canonical choice'' of
identifications is given in \cite{CLJ} (see (2) above).

\item{(6)}{\it infinitesimal $\hb$}\hfill\break
In the framework of nonstandard analysis \cite{AlbevBook} the limit
$\hb\to0$ can be carried out simply by taking $\hb$ literally
infinitesimal. The art, as usual in this theory, is to extract from
the resulting structure the relevant ``standard part''. For the
classical limit the idea is essentially taken from \Thm/T.conv/: the
relevant observables for the classical limit are those, which are
strongly continuous for phase space translations ``on the standard
scale''. Up to corrections of infinitesimal norm this observable
algebra is precisely the algebra $\A_0$ obtained above \cite{IHQ}.
This formulation is perhaps even closer to physical intuition than
the one presented here. However, for the proofs we mostly had to
go back to the standard proofs given in this paper.

\item{(7)}{\it Spin systems}\hfill\break
Of course, one can also consider particles with spin, or other
internal degrees of freedom. How the classical limit on these
internal degrees of freedom is to be taken depends on the physical
question under consideration. For example, in the kinetic theory of
gases, one sometimes leaves these degrees of freedom untouched,
obtaining a theory of classical particles with quantum excitations.
But we can also fix the spin in angular momentum units, which means
that the half-integer labelling the irreducible representation of
$SU_2$ must go to infinity. This limit can be stated exactly along
the lines of this paper, with analogous results. It is essentially
equivalent to a mean-field limit \cite{FGH}. It can also be carried
out for systems of many spins \cite{IMF}, for more general compact
Lie groups \cite{Nick}, and for some quantum groups \cite{FGH}.
For a nonstandard version, see \cite{IHQ}.

\def\refskip{\vskip 8pt plus 2pt}
\let\REF\doref
\Acknow
This paper has grown out of a series of lectures given at the Marc
Kac Seminar in Amsterdam in Summer 1993. The topic of the lectures
was non-commutative large deviation theory, and the classical limit
was included at the request of some members of the seminar, taking me
up on my claim that the techniques I was presenting had applications
to this problem. I would like to thank the members of the seminar,
and in particular the organizers, Hans Maassen and Frank den
Hollander, for the stimulating atmosphere of the seminar.

\REF AHK Albeverio \Jref
    S. Albeverio, R. H\o egh-Krohn
    "Oscillatory integrals and the method of stationary phase in
    infinitely many dimensions, with applications to the classical
    limit of quantum mechanics, I"
    Invent.Math. @40(1977) 59--106

\REF AFHL AlbevBook \Bref
    S. Albeverio, J.E. Fenstad, R. H\o egh-Krohn, T. Lindstr{\o}m
    "Nonstandard methods in stochastic analysis and mathematical
    physics"
    Academic Press, Orlando 1986

\REF Ara Arai \Jref
    T. Arai
    "Some extensions of the semiclassical limit $\hbar\to0$ for
    Wigner functions on phase space"
    J.Math.Phys. @36(1995) 622--630

\REF  BB  Berry \Jref
    M.V. Berry, N.L. Balazs
    "Evolution of semiclassical quantum states in phase space"
    J.Phys. @A12(1979) 625--642

\REF BV Belliss \Jref
     J. Bellissard, M. Vittot
     "Heisenberg's picture and non commutative geometry of the semi-
     classical limit in quantum mechanics"
    Ann.Inst.Henri Poincar\'e @A52(1990) 175--235

\REF BCSS Sirugue \Gref
    Ph. Blanchard, Ph. Combe, M. Sirugue, M. Sirugue-Collin
    "Estimates of quantum deviations from classical mechanics using
    large deviation results"
    \inPr L. Accardi, W. von Waldenfels
    "Quantum probability and applications II"
    Springer Lect.Not.Math. 1136, Berlin 1985

\REF Bop Bopp \Jref
    F. Bopp
    "La m\'echanique quantique est-elle une m\'echanique statistique
    classique particuli\`ere?"
    Ann.Inst. Henri Poincar\'e @15(1956) 81--112

\REF BW PWI \Jref
    T. Br\"ocker, R.F. Werner
    "Mixed states with positive Wigner functions"
    J.Math.Phys. @36(1995) 62--75

\REF Bru Bruer \Jref
    J.T. Bruer
    "The classical limit of quantum theory"
    Synthese @50(1982) 167--212

\REF BS BurdHJ \Jref
    M. Burdick, H.-J. Schmidt
    "On the validity of the WKB approximation"
    J.Phys. A @27(1994) 579--592

\REF Car Cartwright \Jref
    N.D. Cartwright
    "A non-negative Wigner-type distribution"
    Physica @83A(1976) 210--212

\REF Chu Chung \Bref
    K.L. Chung
    "A course in probability theory"
    Academic Press, New York 1974

\REF Dau Daubechies \Jref
    I. Daubechies
    "On the distributions corresponding to bounded
     operators in the Weyl quantization"
    Commun.Math.Phys. @75(1980) 229--238\more; \hfill\break
    \Jref      \sameauthor. {}
    "Continuity statements and counterintuitive examples
     in connection with Weyl quantization"
    J.Math.Phys. @24(1983) 1453--1461

\REF Dav Davies \Bref
     E.B. Davies
     "Quantum theory of open systems"
     Academic Press, London 1976

\REF DGI Esposti \Jref
    M. Degli Esposti, S. Graffi, S. Isola
    "Classical limit of the quantized hyperbolic toral
    automorphisms"
    Commun.Math.Phys. @167(1995) 471--507

\REF DH Duclos \Jref
    P. Duclos, H. Hogreve
    "On the semiclassical localization of quantum probability"
    J.Math.Phys. @34(1993) 1681--1691

\REF Duf Nick \Gref
    N.G. Duffield
    "Classical and thermodynamic limits for generalised quantum spin
    systems"
    Commun.Math.Phys. @127(1990) 27--39

\REF DW LMD   \Jref
    N.G. Duffield,  R.F. Werner
    "Local dynamics of mean-field quantum systems"
    Helv.Phys.Acta @65(1992) 1016--1054

\REF Em1 Emch \Jref
     G.G. Emch
    "Geometric dequantization and the correspondence problem"
     Int.J.Theo.Phys. @22(1983) 397--420

\REF Em2 Emchbook \Bref
    \sameauthor. {}
    "Conceptual and mathematical foundations of 20th-century
    physics"
    North-Holland Mathematics Studies 100,
    Amsterdam 1984

\REF FLM Hajo \Jref
    W. Fischer, H. Leschke, P. M\"uller
    "Dynamics by white noise Hamiltonians"
    Phys.Rev.Lett. @73(1994) 1578--1581

\REF Fr\"o  Froman \Bref
    N. Fr\"oman, P.O. Fr\"oman
    "JWKB Approximation; Contributions to the theory"
    North-Holland, Amsterdam 1965

\REF GW FGH \Gref
    C.-T. Gottstein, R.F. Werner
    "Ground states of the infinite q-deformed Heisenberg
    ferromagnet"
    Preprint, Osnabr\"uck 1994
    archived at {\tt cond-mat/9501123}

\REF Gro Grossman \Jref
   A. Grossmann
   "Parity operator and quantization of delta-functions"
   Commun.Math.Phys. @48(1976) 191--194

\REF Hag Hagedorn \Gref
    G.A. Hagedorn
    "Semiclassical quantum mechanics"
    \more    \Jref part I\eat. {}
    "The $\hbar\to0$ limit for coherent states"
    Commun.Math.Phys. @71(1980) 77--93,
    \more    \Jref part III\eat. {}
    "The large order asymptotics and more general states"
    Ann.Phys. @135(1981) 58--70,
    \more    \Jref part IV\eat. {}
    "Large order asymptotics and more general states in
    more than one dimension"
    \hfill\break
    Ann.Inst.Henri Poincar\'e @A42(1985) 363--374

\REF Hel Helffer \Bref
    B. Helffer
    "Semi-classical analysis for the Schr\"odinger operator and
    applications"
    Springer Lect.Not.Math. 1336, Berlin 1988

\REF Hep Hepp \Jref
    K. Hepp
    "The classical limit for quantum mechanical correlation
     functions"
    Commun.Math.Phys. @35(1974) 265--277.

\REF Hol Holevo \Bref
    A.S. Holevo
    "Probabilistic and statistical aspects of quantum theory"
    North Holland, Amsterdam 1982

\REF Hor Hormander \Bref
    L. H\"ormander
    "The analysis of linear partial differential operators"
    ?? volumes
    Springer, Berlin 1985

\REF Hud Hudson \Jref
    R.L. Hudson
    "When is the Wigner quasi-probability density nonnegative ?"
    Rep.Math.Phys. @6(1974) 249--252

\REF Kat Kato    \Bref
    T. Kato
    "Perturbation theory for linear operators"
    Springer, Berlin, Heidelberg, New York 1984

\REF Kat Boch  \Bref
    Y. Katznelson
    "An introduction to harmonic analysis"
    Wiley\&Sons, New York 1968

\REF Lan Landsman \Jref
    N.P. Landsman
    "Deformations of algebras of observables and the classical limit
    of quantum mechanics"
    Rev.Math.Phys. @5(1993) 775--806

\REF Lie Liebspin \Jref
    E.H. Lieb
    "The classical limit of quantum spin systems"
    Commun.Math.Phys. @62(1973) 327--340

\REF LS Liebcoh \Jref
    E.H. Lieb, J.P. Solovej
    "Quantum coherent operators: a generalization of coherent
    states"
    Lett.Math.Phys. @22(1991) 145--154

\REF Mas Maslov \Bref
    V.P. Maslov, M.V. Fedoriuk
    "Semi-classical approximation in quantum mechanics"
    D. Reidel, Dordrecht 1981

\REF Nar Narcowich \Jref
    F.J. Narcowich
    "Conditions for the convolution of two Wigner distributions to
    be itself a Wigner distribution"
    J.Math.Phys. @29(1988) 2036--2041

\REF Omn Omnes \Gref
    R. Omn\`es
    "Logical reformulation of quantum mechanics"
    \more    \Jref part III\eat. {}
    "Classical limit and irreversibility"
    J.Stat.Phys. @53(1988) 957--975
    \more    \Jref part IV\eat. {}
    "Projectors in semiclassical physics"
    J.Stat.Phys. @57(1989) 357--382

\REF RW IMF  \Jref
    G.A. Raggio, R.F. Werner
    "The Gibbs variational principle for inhomogeneous mean
        field systems"
    Helv.Phys.Acta @64 (1991) 633--667

\REF Ri1 Rieffel \Jref
    M.A. Rieffel
    "Deformation quantization of Heisenberg manifolds"
    Commun.Math.Phys. @122(1989) 531--562

\REF Ri2 RieffAMS \Bref
    M.A. Rieffel
    "Deformation quantization for actions of $\Rl^d$"
    Memoirs of the AMS, \#506,
    Am.Math.Soc., Providence 1993

\REF Ri3 Rieffprep \Gref
    M.A. Rieffel
    "The classical limit of dynamics for spaces quantized by an
    action of $\Rl^d$"
    Preliminary version, March 1995

\REF Rob Robert \Bref
    D. Robert
    "Autour de'l approximation semi-classique"
    Birkh\"auser, Boston 1987

\REF Sch Schiff \Bref
    L.I. Schiff
    "Quantum mechanics"
    McGraw-Hill Kogakusha, Tokyo 1955

\REF Sim Simon    \Jref
    B. Simon
    "The classical limit of quantum partition of functions"
    Commun.Math.Phys. @71(1980) 247--276

\REF Tak Takahashi \Jref
    K. Takahashi
    "Wigner and Husimi functions in quantum mechanics"
    J.Phys.Soc.Jap. @55(1986) 762--779

\REF Tru Aubrey \Jref
    A. Truman
    "Feynman path integrals and quantum mechanics as $\hbar\to0$"
    J.Math.Phys. @17(1976) 1852--1862

\REF Vor Voros \Jref
    A. Voros
    "An algebra of pseudodifferential operators and the asymptotics of
     quantum mechanics"
    J.Funct.Anal. @29(1978) 104--132

\REF We1 QHA \Jref
    R.F. Werner
    "Quantum harmonic analysis on phase space"
    J.Math.Phys. @25(1984) 1404--1411

\REF We2 PHU \Jref
    \sameauthor. {}
    "Physical uniformities on the state space of
        non-relativistic quantum mechanics"
    Found.Phys. @13(1983) 859--881

\REF We3 KAC \Gref
    \sameauthor. {}
    "Mean field quantum systems and large deviations"
    Lectures in the Mark Kac Seminar, Summer 1993;
    Written version in preparation as ``Quantum lattice systems with
    infinite range interactions''

\REF We4 CLJ \Gref
    \sameauthor. {}
    "The classical limit of quantum mechanics as an inductive limit"
    In preparation

\REF We5 CLD \Gref
    \sameauthor. {}
    "The classical limit of quantum mechanics: differentiable
    structure and dynamics"
    In preparation

\REF We6 CLN \Gref
    \sameauthor. {}
    "The classical limit of quantum mechanics: norm limit of states
    and equilibrium states"
    In preparation

\REF WW IHQ \Gref
    R.F. Werner, M.P.H. Wolff
    "Classical mechanics as quantum mechanics with infinitesimal
    $\hbar$"
    Preprint Osnabr\"uck, Oct.~1994,
    archived at {\tt mp\_arc, \# 94-388}

\REF Wig Wigner \Jref
    E.P. Wigner
    "On the quantum correction for thermodynamic equilibrium"
    Phys.Rev. @40(1932) 749--759

\REF WS Wresin \Jref
    W. Wreszinski, G. Scharf
    "On the relation between classical and quantum statistical
    mechanics"
    Commun.Math.Phys. @110(1987) 1--31

\bye